\documentclass[epj]{svjour}
\usepackage{graphicx}
\usepackage{amssymb}
\usepackage{amsmath}
\usepackage{verbatim}
\usepackage{xcolor}
\usepackage{url}
\usepackage{multirow}
\usepackage{booktabs}
\begin{document}
\title{Cosmological constraints on the $R^2$-corrected Appleby-Battye model}
\author{Bruno Ribeiro\inst{1,}\thanks{brunoribeiro@on.br}
\and Armando Bernui\inst{1,}\thanks{bernui@on.br}
\and Marcela Campista\inst{2,}\thanks{marcelacampista@macae.ufrj.br}
%
}                     
%
%
\institute{Observatório Nacional, Rio de Janeiro 20921-400, Brazil \and Universidade Federal do Rio de Janeiro, Macaé 27930-560, Brazil}
\date{Received: date / Revised version: date}
%
\abstract{
Nowadays, efforts are being devoted to the study of alternative cosmological scenarios, in which, modifications of the General Relativity theory have been proposed to explain the late cosmic acceleration without assuming the existence of the dark energy component. 
In this scenario, we investigate the $R^2$-{\it corrected Appleby-Battye model}, or $R^2$-AB model, which consists of an $f(R)$ model with only one extra free parameter $b$, besides the cosmological parameters of the flat-$\Lambda$CDM model: $H_0$ and $\Omega_{m,0}$.
Regarding this model, it was already shown that a positive value for $b$ is required for the model to be consistent with Solar System tests, moreover, the condition for the existence of a de~Sitter state requires $b \ge 1.6$.
To impose observational constraints on the $R^2$-AB model we consider three 
datasets: $31$ $H(z)$ measurements from Cosmic Chronometers (CC), $20$ $[{f\sigma}_{8}](z)$ measurements from Redshift-Space Distortion (RSD), and the most recent type Ia Supernovae (SNe Ia) sample from Pantheon+. 
Next, we perform two diferent analyses: we have cosidered only SNe Ia data and the combined likelihood SNe$+$CC$+$RSD. 
The first one has provided $b=2.28^{+6.52}_{-0.55}$, while the second one $b=2.18^{+5.41}_{-0.55}$. In the first case it was necessary to set the absolute magnitude $M_B = -19.253$ from SH$0$ES collaboration, while in the second we did a marginalization over the Hubble constant $H_0$ in the normalized growth function. 
We have also observed that the $H_0-M_B$ degenerecency was broken by adding CC data to the SNe data. 
Additionally, we perform illustrative analyses that compare this $f(R)$ model with the flat-$\Lambda$CDM model, considering several values of the parameter $b$, for diverse cosmological functions like the Hubble function $H(z)$, the equation of state $w_{\rm eff}(z)$, the parametrized growth rate of cosmic structures $[f \sigma_8](z)$, and $\sigma_8(z)$.
From our results, we conclude that the $R^2$-AB model fits well current observational data, 
although the model parameter $b$ was not unambiguously constrained in the analyses.
}

%
\maketitle
%


\section{Introduction}
\label{intro}
The recognition of the flat-$\Lambda$CDM model as the best model to describe the Universe is almost a consensus, and this is undoubted because, besides explaining the recent phase of accelerated cosmic expansion discovered by Riess et al.~\cite{Riess1998} and Perlmutter et al.~\cite{Perlmutter1999}, this model is in agreement with a plethora of observations, e.g., CMB,  BAO, SNe, gravitational lensing, etc., motivating to call it the concordance cosmological model. 
However, this does not mean that it will remain the concordance model. The large volume of available data and the evolution of statistical procedures have transported cosmology from the era of precision to the era of accuracy. 
Recent results have revealed several tensions at large and small scales, hitherto hidden~\cite{Riess2022,Anchordoqui2021}. These tensions, added to the unknown nature of the dark energy (DE) component, have motivated studies of alternative scenarios (see, e.g.,~\cite{Yang:2021eud,Odintsov:2020qzd,Bernui2023}).

The $\Lambda$CDM-type cosmological models, which include the flat-$\Lambda$CDM case, assume the General Relativity (GR) theory as the metric theory. On the other hand, alternative cosmological models propose to give up this hypothesis, by considering other metric theories as extra dimensions, extra fields, and higher orders corrections, termed modified gravity (MG) theories. In some cases, the cosmic accelerated expansion comes from the gravity model assumed, and the DE is no longer needed.

Generically, an MG theory assumes that the metric theory of the $n+1$ dimensional space-time describing the universe phenomena at quantum and large scales comes from a suitable modification of the GR. This is because at Solar System scale, and also in the distant past $z \gg 1$, GR has passed with honors a set of astrophysical observations confirming its validity. Thus, the first MG theories were proposed to renormalize the GR theory and to obtain a classical theory of gravitation as the low-energy limit of the quantum gravity \cite{Capozziello2011}. 
Recently, some MG theories have been proposed to explain the cosmic acceleration of the universe, this is the case of the so-called $f(R)$ theories~\cite{Clifton2012}.

The $f(R)$ theories are thought as modifications of the GR theory, obtained when we replace the term $R - 2\Lambda$ in the Einstein-Hilbert (EH) Lagrangian by an arbitrary function of the Ricci scalar $R$. This class of theories is conformally equivalent to Einstein's theory, with the addition of an extra degree of freedom in the gravitational sector, the {\em scalaron}, a canonical scalar field whose potential is uniquely determined by the scalar curvature 
$R$~\cite{DeFelice2010,Sotiriou2008}.

The first successful $f(R)$ model has been proposed by Starobinsky in order to explain the primordial inflationary era, with the generic form $R + \epsilon R^2$, $\epsilon > 0$~\cite{Starobinsky1980}, and from then on, several $f(R)$ models have been proposed, considering from simple polynomial laws to more complicated functions of the Ricci scalar (see, e.g., 
Amendola et al.~\cite{Amendola2006}, Hu-Sawicki~\cite{Hu2007}, Starobinsky~\cite{Starobinsky2007}, Appleby-Battye~\cite{Appleby2007}, 
Li-Barrow~\cite{Li2007}, Amendola-Tsujikawa~\cite{Amendola2008}, Tsujikawa~\cite{Tsujikawa2008}, Cognola et al.~\cite{Cognola2008}, 
Linder~\cite{Linder2009}, Elizalde et al.~\cite{Elizalde2011}, Xu-Chen~\cite{Xu2014}, Nautiyal et al.~\cite{Nautiyal2018}, 
Gogoi-Goswami~\cite{Gogoi2020}, and Oikonomou~\cite{Oikonomou2013,Oikonomou2021}).

The $f(R)$ models offer alternative scenarios where the recent cosmic acceleration is an effect of the space-time geometry, instead of an unknown and exotic form of energy. However, most of them are discarded for theoretical reasons, surviving those who reduce to GR in some limit, for this called {\it viable} models~\cite{Amendola2006}. In fact, a critical feature is that, in general, they cannot be naturally incorporated into any high-energy theory, they need a proper fine tuning related to the unbounded growth of the scalaron mass~\cite{Tsujikawa2008}. 
Various approaches have been proposed to mitigate this problem, such as introducing additional terms in the $f(R)$ action, like an $R^2$ term, with a suficiently small coefficient to ensure the existence of the primordial inflation~\cite{DeFelice2010,Starobinsky2007,Appleby2010}. These additional terms are intended to stabilize the scalaron mass and alleviate the fine-tuning requirements, making the models consistent with observations.

Notice, however, that within the class of viable models there is a degeneracy because diverse $f(R)$ models correctly describe the accelerated cosmic expansion as the flat-$\Lambda$CDM does. 
In these cases, one has to go to the perturbative level to decide which model reproduces better the matter clustering in the observed  universe~\cite{Avila2019,Marques2020,deCarvalho2021,Franco23,Oliveira23}. 

It is known that the alternative models have a larger number of parameters when compared to the flat--$\Lambda$CDM, with only one free parameter, and this is the reason why these models undergo a statistically less efficient process of best-fitting cosmological data. However, this is not always true when Bayesian statistical analysis is considered since in Bayesian approach the comparison can vary significantly with the prior choices. 

In this work we shall study the $R^2$-corrected Appleby-Battye (AB) model, proposed in~\cite{Appleby2010}, that for the sake of simplicity, 
it will be denoted throughout the text by $R^2$-AB model.
This model results from the improvement of the original AB model~\cite{Appleby2007}, where a term proportional to $R^2$, 
with a sufficiently small coefficient to ensure the existence of the primordial inflation, was added to solve the {\em weak curvature singularity} 
problem~\cite{DeFelice2010,Appleby2010}, present in a number of $f(R)$ models.
The reasons for analyzing this model are diverse; 
firstly, 
the fact that it passed many important tests
(e.g., classical and semi-classical stability, solar system constraints, correct primordial nucleosynthesis of light elements, and has radiation, matter, and DE  epochs), 
makes this model a viable alternative to explain the current accelerated epoch. 
Besides, there are no analyses of this model involving cosmological data to 
investigate its model parameters. 
Note that, if this model shows good agreement with the observational data, it will provide a geometrical explanation for the accelerated expansion, not being necessary to assume a (non-physical) dark energy component in the universe.

This work is organized as follows. 
In section~\ref{frgravity} we introduce the formal basis for $f(R)$ gravity within the framework of the metric formalism. In section~\ref{frcosmology} a brief description of the $f(R)$ cosmology is presented. Throughout these sections, we highlight important sets of constraints that the $f(R)$ models must obey. Next, in section~\ref{R2model}, we present the main aspects of the $R^2$-AB model and compare it theoretically with the flat-$\Lambda$CDM model. In section~\ref{dataset} we provide two datasets and details on their compilations, measurements, surveys, and cosmological tracers. Finally, we describe the statistical methodology of the analysis performed and show our results in section~\ref{results}, as well as address our conclusions in section~\ref{conclusions}. 
Additionally, we provide some plots in appendix~\ref{apen} emphasizing the role of the free parameter of the $R^2$-AB model versus the flat-$\Lambda$CDM model, and highlight some basic differences between the $R^2$-AB model and the Hu-Sawicki and Starobinsky models in the appendix~\ref{apen2}.

Through this work we assume natural units in which $\hbar=c=1$. Greek symbols range from $0$ to $i$, with $0$ being the cosmic time and $i=1,2,3$ the three-dimensional space. The covariant derivative is denoted by $\nabla_\mu$, and $\Box \equiv g^{\mu\nu}\nabla_\mu\nabla_\nu$ is the d'Alembertian operator.


\section{\label{frgravity}$f(R)$ gravity in brief: the state of the art}

The modified Einstein-Hilbert (EH) action for $f(R)$ gravity~\cite{Capozziello2011,Clifton2012,DeFelice2010,Sotiriou2008,Faraoni2010,Papantonopoulos2014} is 
\begin{equation}\label{acao}
    S = \int d^4x\sqrt{-g}\left[\frac{M^2_{\mbox{\footnotesize P}}}{2}f(R) + \mathcal{L}_M\right],
\end{equation}
where $M_{\mbox{\footnotesize P}}^2 \equiv (8\pi G)^{-1}$ is the Planck mass and $\mathcal{L}_M$ is the matter Lagrangian describing the material content. Varying the modified EH action with respect to the metric variable, we get the following field equations 
\begin{equation}\label{eqcampo}
    f^\prime(R)R_{\mu\nu} - \frac{1}{2}g_{\mu\nu}f(R) - \left(\nabla_\mu\nabla_\nu - g_{\mu\nu}\,\Box\right)f^\prime(R)=\frac{T_{\mu\nu}}{M^2_{\mbox{\footnotesize P}}}\,,
\end{equation}
and the trace equation 
\begin{equation}\label{traco}
    R f^\prime(R) - 2 f(R) + 3\,\Box 
    f^\prime(R) = \frac{T}{M^2_{\mbox{\footnotesize P}}}\,,
\end{equation}
where $f^\prime(R)\equiv df(R)/dR$ and $T\equiv T^\mu_\mu$ is the trace of the stress-energy tensor. 
In the absence of matter, the exact solution of eq.~(\ref{eqcampo}) is 
\begin{equation}\label{vacuo}
    R f^\prime(R) - 2f(R)=0\,,
\end{equation}
where the positive real roots of this equation yield the well-known de Sitter vacuum solutions. 
These solutions form the basis for the description of the early and late acceleration phases of the universe. 
However, as pointed out in~\cite{Muller1987}, there are acceptable solutions if 
\begin{equation}
    \left.\frac{f^\prime(R)}{f^{\prime\prime}(R)}\right|_{R=R_*} > R_* \,,
\end{equation}
where $f^{\prime\prime}(R)\equiv d^2f(R)/{dR}^2$, and $R_*$ is a positive real root of  eq.~(\ref{vacuo}). 
In fact, the $f(R)$ function characterizing the MG model is not arbitrary. 
Instead, it must satisfy a set of rules to ensure both theoretical consistency and phenomenological viability of the model.
\begin{itemize}
\item[{\em (i)}] be stable in 
the interval of $R$ of cosmological interest, i.e.,
\begin{equation}\label{i}
f^\prime(R)>0\,, \ f^{\prime\prime}(R)>0;
\end{equation}
\item[{\em (ii)}] have a stable Newtonian limit, i.e., 
\begin{equation}
\left|f(R) - R\right|\ll R\,,\ \left|f^\prime(R) - 1\right| \ll 1\,, \ Rf^{\prime\prime}(R)\ll 1\,,
\end{equation}
for $R \gg R_0$, where $R_0$ is the curvature scalar today;
\item[{\em (iii)}] be indistinguishable from GR at the current level of accuracy of laboratory experiments and tests of the Solar System phenomena involving gravity;
\item[{\em (vi)}] recover the GR in the high-curvature regime, i.e.,
\begin{equation}\label{RggR0}
\lim_{R \gg R_0} f(R)= R - 2\Lambda \,.
\end{equation}
This constraint also means that $f^\prime(\infty) = 1$, implying that $0 < f^\prime(R) < 1$ for all $R$; and
\item[{\em (v)}] have a stable (or metastable) asymptotic de Sitter future.

\end{itemize}
In the first condition {\em (i)}, $f^\prime(R)>0$, ensures that gravity be attractive (i.e., the {\em gravitons} are not ghosts)~\cite{Nunez2004,Krause2006,Himmetoglu2009,Deruelle2011}, while the second, $f^{\prime\prime}(R)>0$, avoids the {\em scalaron} from becoming a {\em tachyon}, in the high-curvature regime~\cite{Dolgov2003,Olmo2005,Faraoni2006}. 
Next, given the success of the Newtonian theory in explaining the observed non-homogeneities at small scale and compact objects, it is necessary to impose that $f(R)$ model recovers this theory when $R \gg R_0$ at these scales, as indicated at {\em (ii)}. 
The GR theory is very well-tested in the laboratory and in the Solar System, where no significant deviations from the theory have been observed to date. 
Besides, observational data from the cosmic microwave background (CMB) regarding processes of the early universe strongly agree with the robust predictions of the concordance flat-$\Lambda$CDM model, such as the  big bang nucleosynthesis (BBN), a primitive radiation-dominated age, and another middle age matter-dominated. 
Furthermore, we need the $f(R)$ theory to recover the GR in the weak-field regime and in the distant past, as determined at {\em (iii)} and {\em (iv)}. Finally, for a description of the current DE-dominated era, the $f(R)$ model needs to have a stable (or metastable) de~Sitter phase. 
As we shall see in the next section, new constraints on the $f(R)$ will be placed exploring the cosmological viability of the model.


\section{\label{frcosmology}$f(R)$ cosmology}

In this section, we briefly discuss the cosmic dynamics in the $f(R)$ cosmology, 
finding the evolution equations in the background and perturbative levels for a generic $f(R)$ model.

\subsection{Cosmological background}

In order to derive the dynamics of the cosmological background in the $f(R)$, let us first consider the FLRW metric, 
describing a statistically homogeneous and isotropic universe, given by
\begin{equation}\label{metrica}
{ds}^2= - {dt}^2 + a^2(t)\left(\frac{{dr}^2}{1-Kr^2} + r^2{d\Omega}^2\right)\,, 
\end{equation}
where $K=0,\pm 1$ and $a(t)$ is the scale factor. 
Assuming a flat spatial section ($K=0$) and replacing eq.~(\ref{metrica}) into eq.~(\ref{eqcampo}), 
we get the Friedmann equations
\begin{equation}\label{frd1}
   3H^2 = \frac{\rho_M + \rho_{DE}}{M^2_{\mbox{\footnotesize P}}},
\end{equation}
\begin{equation}\label{frd2}
    2\dot{H} + 3H^2= -\frac{P_M + P_{DE}}{M^2_{\mbox{\footnotesize P}}} \,,
\end{equation}
where the overdot denotes differentiation with respect to cosmic time, $t$, and
\begin{equation}\label{de1}
\begin{aligned}
    \frac{\rho_{DE}}{M^2_{\mbox{\footnotesize P}}} \equiv & \frac{1}{2}\left[Rf^\prime(R) - f(R)\right] - 3H\dot{f}^\prime(R) \\
    & + 3H^2\left[1 - f^\prime(R)\right]\,,
\end{aligned}
\end{equation}
\begin{equation}\label{de2}
\begin{aligned}
    \frac{P_{DE}}{M^2_{\mbox{\footnotesize P}}} \equiv & - \frac{1}{2}\left[Rf^\prime(R) - f(R)\right] + 2H\dot{f}^\prime(R) \\
    &  - (2\dot{H} + 3H^2)\left[1 - f^\prime(R)\right] + \ddot{f}^\prime(R).
\end{aligned}
\end{equation}
Dividing eq.~(\ref{de2}) by eq.~(\ref{de1}) we get the equation of state for the DE gravitational component in 
the $f(R)$ theory, $w_{DE} = P_{DE}/\rho_{DE}$, where $w_{DE}$ is its state parameter; this DE component obeys the continuity equation 
\begin{equation}
    \dot{\rho}_{DE} + 3H\left(\rho_{DE} + P_{DE}\right) = 0 \,.
\end{equation}
In particular, in the GR case, where $f(R) \equiv R - 2\Lambda$, 
the eqs.~(\ref{de1}) and~(\ref{de2}) reduce to the perfect fluid equations, with $P_{DE}=-\rho_{DE}$, hence $w_{DE} \equiv w_{\Lambda}= - 1$. 

Finally, the deceleration parameter is given by
\begin{equation}
q\left(t\right) \equiv - \frac{\ddot{a}}{aH^2} = - 1 - \frac{\dot{H}}{H^2}.
\end{equation}
The $f(R)$ is expected to describe the current accelerated epoch, $q(t_0)<0$, as well as reduce to GR at the high-curvature regime (e.g., as in the solar system neighborhood), and for high-redshift data $z \gg 1$.

The effect of a general function $f(R)$ on cosmological dynamics can be analyzed from a geometric point of view. It is useful to define a new set of functions $(p,s)$ and to study the solutions of $p = p(s)$ on this plane. 
Using the eqs.~(\ref{frd1}), (\ref{frd2}), (\ref{de1}), and (\ref{de2}) we define 
\begin{equation}
p \equiv \frac{Rf^{\prime\prime}(R)}{f^\prime(R)}\, , \ 
s \equiv -\frac{Rf^\prime(R)}{f(R)}\, .
\end{equation}
According to~\cite{Amendola2006}, some rules\footnote{since the newly established rules are constraints imposed on $f(R)$, we adopt the numbering sequence, as presented in section \ref{frgravity}.} 
are established for a viable cosmic dynamics, namely: 
\begin{itemize}
\item[{\em (vi)}] A matter-dominated middle epoch, necessary for structure formation, is achieved if 
\begin{equation}\label{vi}
p(s) \approx +0\,, \ \frac{dp(s)}{ds} > -1 \,,
\end{equation}
at $s = -1$; and

\item[{\em (vii)}] The matter-dominated phase, will be followed by a late accelerated phase, only if 
\begin{equation}\label{viia}
0 \leq p(s) \leq 1 \,,
\end{equation}
at $s=-2$, or
\begin{equation}\label{viib}
p(s) = - s - 1\,,\ \frac{\sqrt{3} - 1}{2} < p(s) \leq 1\,,\ \frac{dp(s)}{ds}<-1 \,.
\end{equation}
\end{itemize}
The $\Lambda$CDM model correspond to the straight line $p=0$, $\forall s$, on the plane $(p,s)$. Since the viable $f(R)$ models are those that approach the $\Lambda$CDM at high-curvature (i.e., when $R\gg R_0$), $p(s)$ needs to be close to zero during the matter-driven epoch, as indicated in the first of the conditions in the eq.~(\ref{vi}). 
The second condition, in turn, 
is required to connect the matter-dominated epoch and the current DE-dominated epoch. On the other hand, the constraints in the eq.~(\ref{viia}) indicate a purely 
de Sitter vacuum, $w_{DE}=-1$, while the eqs.~(\ref{viib}) corresponds to a non-phantom attractor, $w_{DE}>-1$. 
Regarding the conditions (\ref{viia}) and (\ref{viib}), the $f(R)$ model needs to satisfy only one of them.

\subsection{Cosmological perturbations}

Perturbations around the FLRW background produced in the primordial universe explain the relevant part of the CMB power spectrum~\cite{Hu2002}. 
When crossing the horizon at the inflationary period, such fluctuations were frozen, becoming the primordial seeds to the growth of the structures 
on large scales after their re-enter. 
At the linear level, and using the Newtonian gauge, the line element can be written as~\cite{Sasaki1984,Mukhanov1992} 
\begin{equation}
    {ds}^2=-\left(1 + 2\Psi\right){dt}^2 + a^2\left(1 - 2\Phi\right)\delta_{ij}{dx}^i{dx}^j,
\end{equation}
where
\begin{equation}
    \delta_{ij} = \left\{
    \begin{array}{ll}
        1, & \text{for} \ i = j, \\
        0, & \text{for} \ i \ne j,
    \end{array}
    \right.
\end{equation}
is the Kronecker delta function, and $|\Psi|, |\Phi| \ll 1$ are small functions of the three-space, $x^i$, and cosmic time, $t$, termed the Bardeen potentials~\cite{Bardeen1882}.

The space-time metric fluctuations give rise to fluctuations in the material content that evolve via gravitational instability~\cite{Mukhanov1992,Bardeen1882,peebles1967,zel1970}, and can be described by the perturbed components of the stress-energy tensor, 
\begin{eqnarray}
    T^0_0 &=& -\left(1 + \delta\right)  \,,  \\
    T^i_0 &=& -\left(\rho + P\right)v^i \,, \\
    T^0_j &=& \left(\rho + P\right)v_j  \,,  \\
    T^i_j &=& \left(P + \delta P\right){\delta}^i_j + P{\Pi}^i_j \,,
\end{eqnarray}
where $\delta\equiv\delta\rho/\rho$ is the matter density contrast, that, like the four-velocities ($v^i$), pressure fluctuations ($\delta P$) and the {\em anisotropic stresses}, depend on the space-time coordinates. 
In the absence of {\em anisotropic stresses},  ${\Pi}^i_j=0$, one has  $\Psi=\Phi$~\cite{Mukhanov1992}.

For small scale fluctuations~($k^2\gg a^2H^2$) of a matter-dominated fluid, $\delta = \delta_m$, one arrives to the equation~\cite{Tsujikawa2007}
\begin{equation}\label{contraste}
\Ddot{\delta}_m + 2H\dot{\delta}_m - 4\pi G_{\rm eff}\rho_m\delta_m = 0 \,,
\end{equation}
where 
\begin{equation}\label{cge}
    G_{\rm eff}\left(a,k\right) \equiv \frac{G}{f^\prime (R)}\frac{1 + 4\,p\left(\frac{k^2}{a^2R}\right)}{1 + 3\,p\left(\frac{k^2}{a^2R}\right)}\,,
\end{equation}
is called the {\em effective gravitational constant}, that depends on the scale factor $a$ and on the scale $k$. 

An observable quantity useful to compare a model with current observations is the growth rate of 
cosmic structures (or simply growth function), $f_g(a)$, defined by~\cite{Strauss1995}
\begin{equation}\label{cresc1}
f_g(a) \equiv \frac{d\ln{\delta_m\left(a\right)}}{d\ln {a}} \,.
\end{equation}
A good approximation for the growth function is given by $f_g(z) \simeq \Omega_m^\gamma(z)$, where $\gamma(z)$ is the growth index~\cite{Wang1998}. 
In DE models based on GR theory $\gamma(z)$ is approximated by the constant value $\gamma \simeq 3\,(1 - w_{DE})/(5 - 6w_{DE})$~\cite{Linder2007}. 
Thus, the $\Lambda$CDM model corresponds to $\gamma \simeq 6/11$. 

The matter fluctuations amplitude, $\sigma_\mathcal{R}(a)$, in turn, are directly observed from the CMB power spectrum and is given 
by~\cite{Nesseris2017}
\begin{equation}\label{sig8}
\sigma_\mathcal{R}\left(a\right) = \sigma_{\mathcal{R},0}\left[\frac{\delta_m\left(a\right)}{\Bar{\delta}_m\left(1\right)}\right] \,,
\end{equation}
where $\sigma_{\mathcal{R},0}$ corresponds to the amplitude at $z=0$ where $\mathcal{R}$ represents a physical scale. 
The constant $\Bar{\delta}_m(1)$ is a Planck normalization factor so that at $z=0$, 
in the $\Lambda$CDM model, 
$\sigma_{\mathcal{R}}(0)\equiv\sigma_{\mathcal{R}, 0}$. 
It is common to perform the measurements of several cosmological tracers at the physical scale\footnote{the Hubble constant $H_0$ is usually written as $H_0 \equiv 100 \,h$ km/s/Mpc, where $h$ is a real adimensional number} $\mathcal{R}=8$~Mpc/h. 
Thus the product between $f_g(a)$ and $\sigma_8(a)$ results in 
\begin{equation}\label{fsig8}
[f{\sigma_8}](a) =  \frac{\sigma_{8,0}}{\Bar{\delta}_m(1)}\left[\frac{d\,\delta_m(a)}{d\ln{a}}\right] \,,
\end{equation}
which measures the matter density perturbation rate at the physical scale of $8$~Mpc$/h$. This combination is used more often than simply $f_g(a)$ to derive constraints for theoretical model parameters because of data availability. 


\section{\label{R2model}$R^2$-corrected AB model}

The first $f(R)$ model proposed by Appleby and Battye in~\cite{Appleby2007} is a two-parameter model\footnote{for simplicity we are omitting the cosmological parameters $H_0$ and $\Omega_{m,0}$.}:
\begin{eqnarray}\label{AB}
f^{\,I}_{AB}(R) \equiv \frac{R}{2} + \frac{\epsilon_{AB}}{2}\ln{\left[\frac{\cosh{\left(\frac{R}{\epsilon_{AB}}-b\right)}}{\cosh b}\right]} \,,
\end{eqnarray}
which at large $R$ mimics the GR theory with a non-true cosmological constant. The free model parameters are $\epsilon_{AB}$ and $b$.
 %
 %
It is worth mentioning that some $f(R)$ models well established in the literature, like Hu-Sawicki~\cite{Hu2007} and Starobinsky~\cite{Starobinsky2007} models, present three parameters.



A sudden {\em weak curvature singularity} is known to form genetically when $f^{\prime\prime}(R)$ becomes zero for some finite value of $R$, such that the condition $f^{\prime\prime}(R)>0$ is marginally violated, leading to two more elementary problems: both an unbounded growth of the {\em scalaron} mass and an undesired overabundance of this particle at high-curvatures. 
In other words, it means that the {\em scalaron} can behave like a {\em tachyon} and that the amplitude of the oscillations in $R$ grows indefinitely when $R \gg R_{vac}$~\cite{Appleby2010,Appleby2008}. 
For the AB model in eq.~(\ref{AB}) and many others discussed in the literature, this occurs for $R$ within the range of cosmological interest, pointing out an incompleteness of these models. 
However, it was observed that the simple addition of a term proportional to $R^2$ in the $f(R)$, with a sufficiently small coefficient to ensure the existence of the primordial inflation, solves this type of singularity~\cite{DeFelice2010,Starobinsky2007}.

In this way it was proposed the $R^2$-corrected Appleby-Battye, or $R^2$-AB, model~\cite{Appleby2010} 
\begin{equation}\label{abcorrigido}
f_{AB}(R) \equiv f^{\,I}_{AB}(R) + \frac{R^2}{6\mathcal{M}^2} \,,
\end{equation}
where $\mathcal{M}$ characterizes a mass scale coinciding with the {\em scalaron} rest-mass whenever low curvature modifications to GR can be neglected.
The above model is equivalent to the improved $gR^2$-AB model setting $g=1/2$, whose the main cosmological constraints have been reported in refs.~\cite{Motohashi2012,Motohashi2014}. 
Constraints from the early universe imply $\mathcal{M} \approx 1.2 \times 10^{-5}M_{\mbox{\footnotesize P}}$ at the end of inflation, while a stable de Sitter vacuum requires
\begin{equation}
    \frac{1}{4} + \frac{0.28}{(b - 0.46)^{0.81}} \leq g \leq \frac{1}{2}\,,
\end{equation}
\begin{eqnarray}
\epsilon_{AB} = \frac{R_{vac}}{2g[b 
+ \ln\,(\,2\cosh b \,)]} \label{vinc} \,,
\end{eqnarray}
where $R_{vac} \equiv 12H_0^2$. 
In our analyses, the $R^2$-AB model we are studying 
considers: $g=1/2$, hence $b \geq 1.6$ and $\epsilon_{AB} =\frac{R_{vac}}{b + \ln{(2\cosh{b})}}$, necessary 
to reproduce the current accelerated expansion of the universe~\cite{Appleby2010,Motohashi2012,Motohashi2014}. 
Since the equation~(\ref{vinc}) relates $\epsilon_{AB}$ to $b$, this model has only one free parameter, namely, $b$ (in fact, one more parameter than the $\Lambda$CDM model).

The {\em scalaron} mass, $\mathcal{M}_s(R)$, is given by
\begin{equation}
\mathcal{M}_s^2\left(R\right) \equiv \frac{f^\prime\left(R\right) - Rf^{\prime\prime}\left(R\right)}{3f^{\prime\prime}\left(R\right)},
\end{equation}
so that, at high-curvature regime ($R \gg R_0$), we must have $\mathcal{M}_s=1/\sqrt{3f^{\prime\prime}(R)}$~\cite{DeFelice2010}. 
Then, one can verify that the upper limit for the {\em scalaron} mass, $\mathcal{M}_s \leq \mathcal{M}$, was made possible by adding the Starobinsky-like term $R^2/6\mathcal{M}^2$.


Assuming that the universe is filled with a fluid, consisting of ordinary matter and radiation, represented by subscript index $m$ and $r$, respectively, and a dark energy component, by subscript index DE, the background dynamical equations are given by:
\begin{equation}\label{H_function}
    3H^2=\frac{\rho_m + \rho_r + \rho_{DE}}{M^2_{\mbox{\footnotesize P}}} \,,
\end{equation}
\begin{equation}\label{H_function2}
    2\dot{H} + 3H^2 = -\frac{P_m + P_r + P_{DE}}{M^2_{\mbox{\footnotesize P}}} \,.
\end{equation}
Previously, it was shown by eqs.~(\ref{de1}) and~(\ref{de2}), that the dark energy component has a purely geometrical origin. For the model (\ref{abcorrigido}):
\begin{equation}\label{denDE}
\begin{aligned}
    \frac{\rho_{DE}}{M^2_{\mbox{\footnotesize P}}} \equiv & - \left(\frac{1}{3\mathcal{M}^2} + \frac{\text{sech}^2{y}}{2\epsilon_{AB}}\right)3H\dot{R} \\ 
    & - \frac{\epsilon_{AB}}{4}\ln{\left(\frac{\cosh{y}}{\cosh{b}}\right)} + \frac{R}{4} - \frac{R^2}{12\mathcal{M}^2} \\
    & + \left[\frac{R}{\mathcal{M}^2} + \frac{3(\tanh{y} - 1)}{2}\right](\dot{H} + H^2) \,,
\end{aligned}
\end{equation}
\begin{equation}\label{preDE}
\begin{aligned}
    \frac{\rho_{DE}}{M^2_{\mbox{\footnotesize P}}} \equiv & \left(\frac{1}{3\mathcal{M}^2} + \frac{\text{sech}^2{y}}{2\epsilon_{AB}}\right)\xi_R + \frac{R^2}{12\mathcal{M}^2} - \frac{R}{4} \\ 
    & + \frac{\epsilon_{AB}}{4}\ln{\left(\frac{\cosh{y}}{\cosh{b}}\right)} - \frac{\dot{R}^2\tanh{y}\,\text{sech}^2{y}}{\epsilon_{AB}^2} \\
    & - \left[\frac{R}{3\mathcal{M}^2} + \frac{(\tanh{y} - 1)}{2}\right]\xi_H \,,
\end{aligned}
\end{equation}
where we define the auxiliary variables $y \equiv R/\epsilon_{AB} - b$, $\xi_R\equiv \ddot{R} + 2H\dot{R}$, and $\xi_H \equiv \dot{H} + 3H^2$. Through the domain eras of radiation and early matter, when $R_{vac}\ll R \ll \mathcal{M}^2$, the expressions in~(\ref{denDE}) and~(\ref{preDE}) can be approximated, respectively, by
\begin{equation}\label{densDE2}
\begin{aligned}
    \frac{\rho_{DE}}{M^2_{\mbox{\footnotesize P}}} \simeq & \frac{R_{vac}}{4} + \frac{1}{\mathcal{M}^2}\left[(\dot{H} + H^2)R - H\dot{R} - \frac{R}{12}\right] \\
    & - \frac{1}{e^{2y}}\left[\frac{6H\dot{R}}{\epsilon_{AB}} + 3(\dot{H} + H^2) + \frac{\epsilon_{AB}}{4}\right],
\end{aligned}
\end{equation}
\begin{equation}\label{preDE2}
\begin{aligned}
    \frac{P_{DE}}{M^2_{\mbox{\footnotesize P}}} \simeq & - \frac{R_{vac}}{4} + \frac{1}{\mathcal{M}^2}\left[\frac{(\xi_R - \xi_HR)}{3} + \frac{R^2}{12}\right] \\
    & - \frac{1}{e^{2y}}\left[\frac{2\xi_R}{\epsilon_{AB}} + \xi_H - \left(\frac{2\dot{R}}{\epsilon_{AB}}\right)^2 + \frac{\epsilon_{AB}}{4}\right].
\end{aligned}
\end{equation}
Since $R$ and $\mathcal{M}$ are both very large in these epochs, we can expect a strong suppression of the last two terms in eqs.~(\ref{densDE2}) and~(\ref{preDE2}), and, consequently, to obtain the state equation $w_{DE}\simeq -1$. 
It means that $R^2$-AB model recovers GR at high-$z$. On the other hand, at later times, when $R\sim R_{vac}$, the {\em scalaron} mass becomes small and significant deviations from $w_{DE}=-1$ must be expected.
\begin{figure}
    \centering
    \includegraphics[scale=0.5]{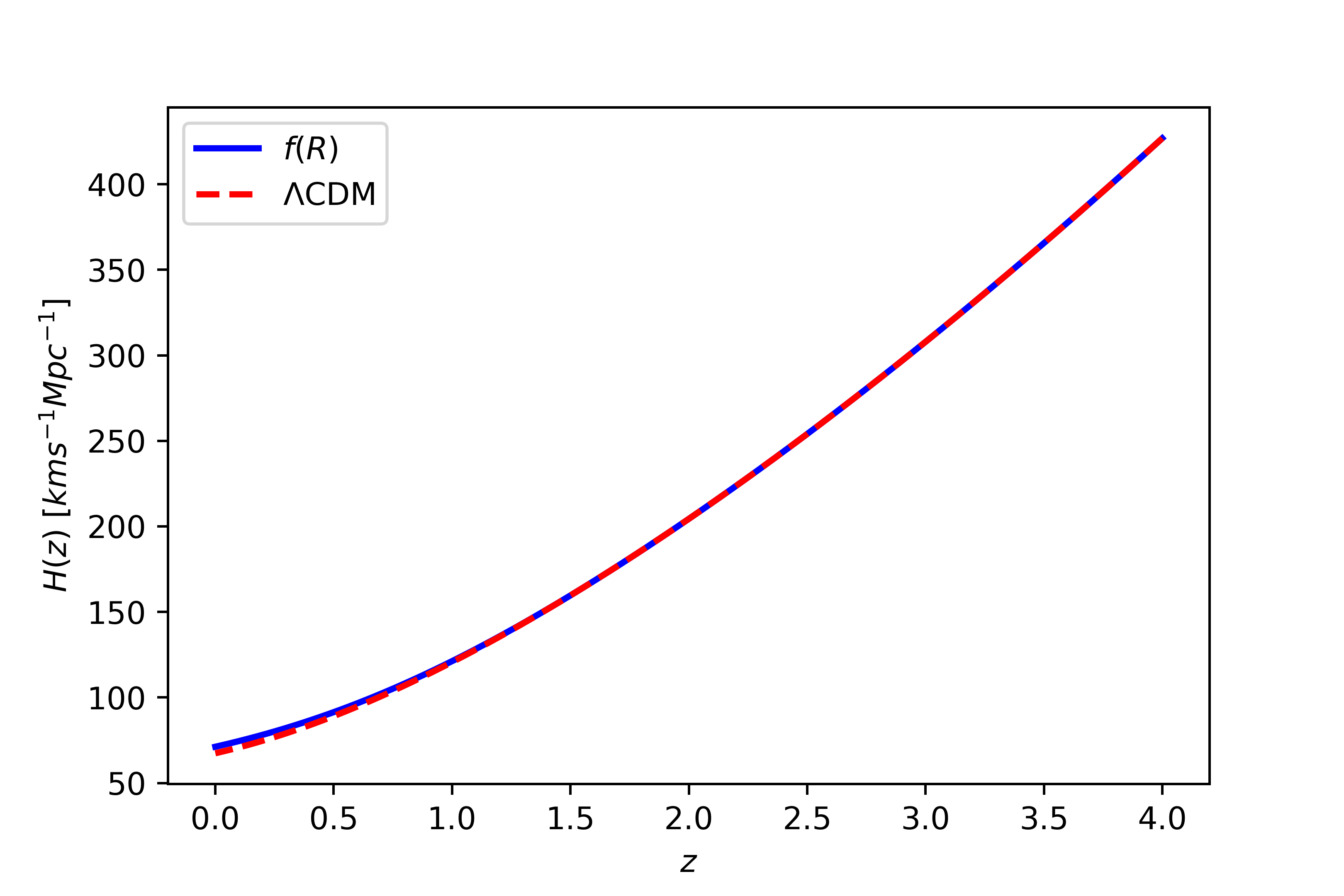}
    \caption{Plot of $H(z)$ functions for the $R^2$-AB and flat-$\Lambda$CDM models. 
    We have used the Planck Collaboration best-fit data: 
    $H_0=67.4$ km/s/Mpc and $\Omega_{m,0}=0.315$~\cite{Planck1}; 
    in addition we set $b=2$ and a scale mass such that $\Delta=10^{-7}$.}
    \label{Hz}
    \end{figure}
\begin{figure}
    \centering
    \includegraphics[scale=0.5]{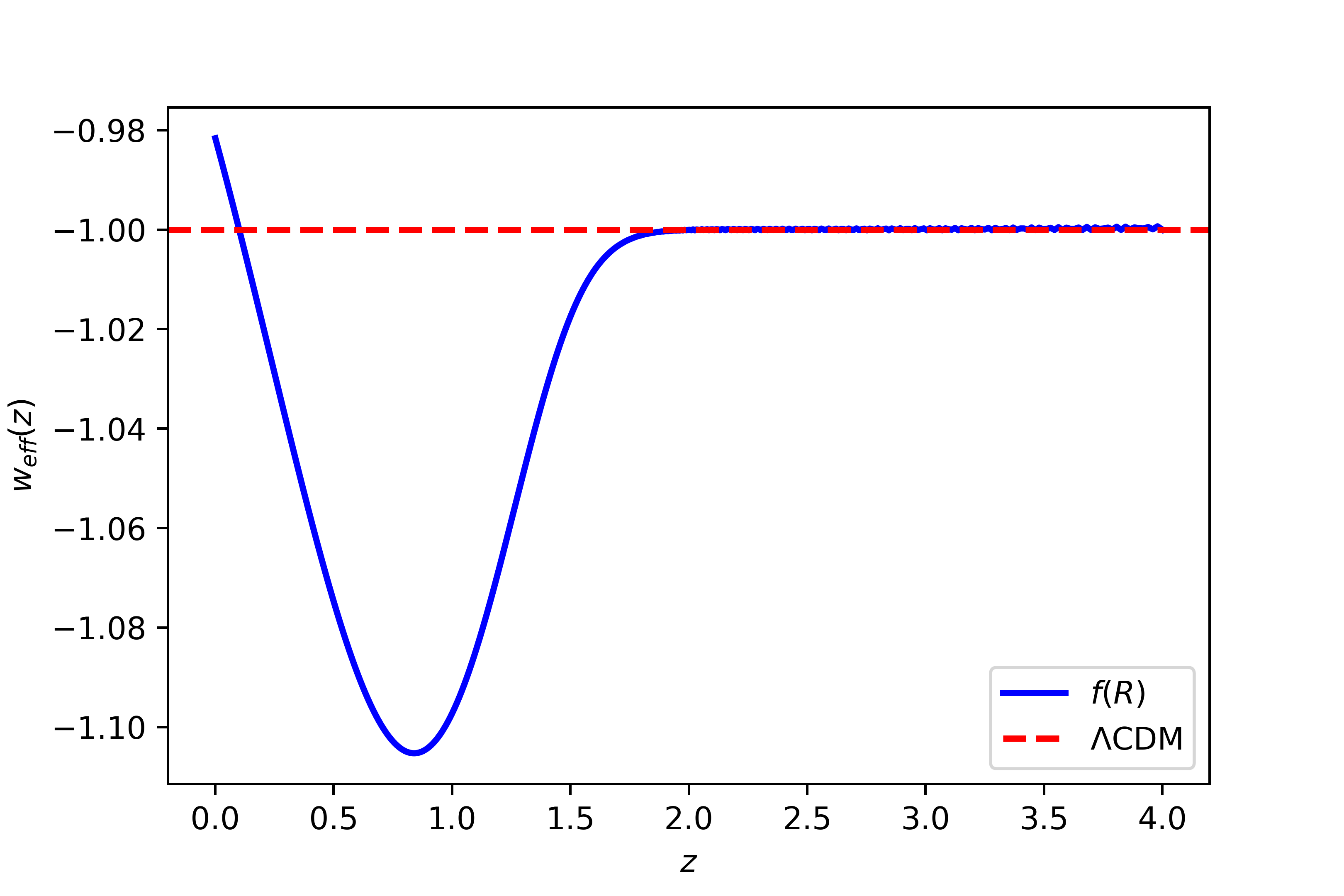}
    \caption{Comparison plot of $w_{\rm eff}(z)$ from the $R^2$-AB model and $w_{\Lambda}=-1$ from fiducial $\Lambda$CDM. 
    We have used $H_0=67.4$ km/s/Mpc, 
    $\Omega_{m,0}=0.315$, $b=2$, and $\Delta=10^{-7}$.}
    \label{wz}
\end{figure}

We now discuss the initial conditions and the most convenient mass scale for the curvature range. 
Firstly, we fix $z_i=4$ as the initial {\em redshift} in our analyses, because we can safely assume that the $\Lambda$CDM describes well the observed universe at this epoch. 
In this way we will have the three necessary initial values $H_i\equiv H_{GR}(t_i)$, $\dot{H}_i\equiv \dot{H}_{GR}(t_i)$ and $\ddot{H}_i\equiv \ddot{H}_{GR}(t_i)$, where $t_i\equiv t(z_i)$. Then, for growing modes in the perturbation level, we assume initial conditions such that $\delta_m(a_i)=a_i$ and $d\,\delta_m(a_i)/da=1$ for $a_i\ll1$~\cite{Nesseris2017}. As suggested in~\cite{Appleby2010}, we can set a mass scale such that $\Delta \equiv \epsilon_{AB}/\mathcal{M}^2=10^{-7}$. 
\begin{figure}
    \centering
    \includegraphics[scale=0.5]{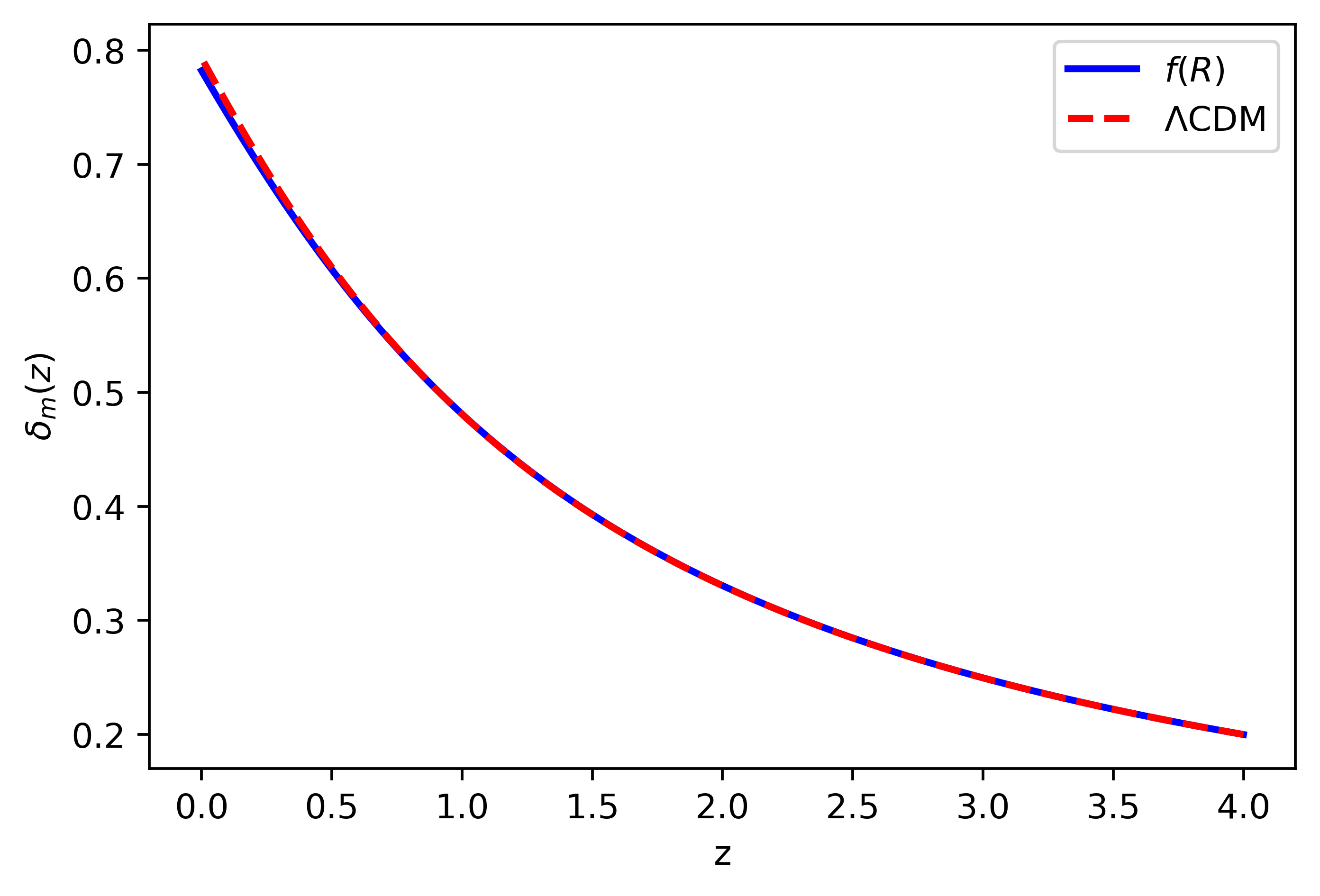}
    \caption{Evolution of the matter density fluctuations, $\delta_m(z)$, for both $R^2$-AB and flat-$\Lambda$CDM models at scale $k=0.125$~Mpc$^{-1}$. 
    We set $\Omega_{m,0}=0.315$, $b=2$, and $\Delta=10^{-7}$. It is worth mentioning that $\delta_m(z)$ is independent on $H_0$.}
    \label{deltaz}
    \end{figure}
\begin{figure}
    \centering
    \includegraphics[scale=0.5]{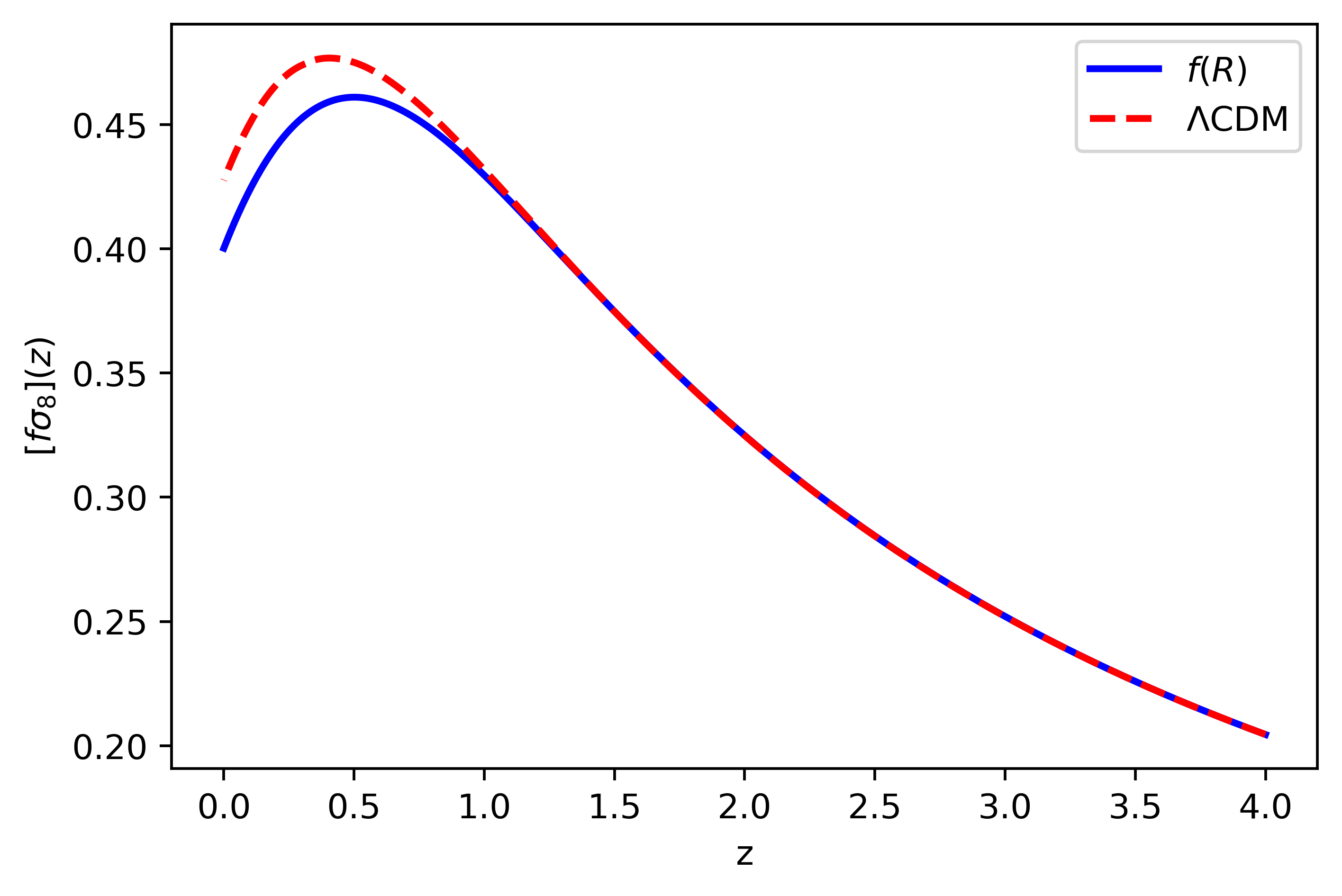}
    \caption{Evolution of $[f\sigma_8](z)$ for $R^2$-AB and flat-$\Lambda$CDM models at the scale $k = 0.125$~Mpc$^{-1}$. 
    In this plot we have used $\Omega_{m,0}=0.315$, $b=2$, $\sigma_{8,0}=0.811$, and $\Delta=10^{-7}$. 
    Like $\delta_m(z)$, $[f\sigma_8](z)$ is also independent of 
    $H_0$.}
    \label{fsig8z}
\end{figure}

In figure~\ref{Hz}, we present the $H(z)$ function , given by eqs.~(\ref{H_function}) and~(\ref{H_function2}), for the $R^2$-AB model.
Additionally, for comparison purposes, we include the fiducial $H(z)$ from the $\Lambda$CDM. Afterward, in figure~\ref{wz}, we show the behavior of the state parameter, $w_{DE}(z)$, where, it is possible to note that the $R^2$-AB model does not reproduces the standard $\Lambda$CDM history around the interval $z<2$. 
This situation changes very little as we vary the parameters involved $H_0$, $\Omega_{m,0}$, and $b$. We can also observe that as we increase the value of $b$, the solution from $f(R)$ approaches that one of $\Lambda$CDM model (see appendix~\ref{apen}). 
In order to understand this behavior, let us consider the limit case $b \gg 1$, or equivalently, $y \gg 1$. In this case, $R^2$-AB model defined in 
eq.~(\ref{abcorrigido}) can be expanded as
\begin{equation}\label{abapprox}
    f(R) \simeq R - \frac{R_{vac}}{2} + \Gamma(R) \,,
\end{equation}
where $\Gamma(R)\equiv (\epsilon_{AB}/2)e^{-2y} + R^2/6\mathcal{M}^2$ and $\Gamma(R) \ll R$. 
Then, for $b$ sufficiently large, the eq.~(\ref{abapprox}) can be written as $f(R) \simeq R - 2\Lambda$, where $\Lambda \equiv R_{vac}/4$. 
In this context, $\Lambda$ is said to be a non-true cosmological constant, as it assumes a geometric (i.e., gravitational) role rather than be part of the material content. 
Another possibility to recover RG is reached by setting $R \gg R_{vac}$, as mentioned in eq.~(\ref{RggR0}).

As commented at the beginning of this section, Hu-Sawicki and Starobinky $f(R)$ models belong to the class of the {\em viable} models. 
Note that these models, as well as $R^2$-AB, present a sudden weak singularity at $R \rightarrow \infty$. 
Both versions of the AB models, i.e., the corrected and uncorrected ones, have been studied in the literature~\cite{Starobinsky2007,Hu2007,Appleby2010,Motohashi2010,Motohashi2013}. The main differences of the cosmological evolution in R$^2$-AB model from Hu-Sawicki and Starobinsky models are discussed in appendix~\ref{apen2}.

Due to degeneracy at the background level, we must look for new cosmological tracers to obtain the best fit for the model parameters and their uncertainties. 
Our first choice is to consider the cosmological perturbations through the matter contrast, $\delta_m(z)$, and the parametrized growth rate of cosmic structures, $[f\sigma_8](z)$. Following this goal, we obtain the eq.~(\ref{contraste}) by  solving eq.~(\ref{fsig8}) for both $f(R)$ and flat-$\Lambda$CDM (for which $G_{\rm eff}\equiv G$) models. 
It is worth mentioning that in MG the structure formation depends on scale $k$ through the effective gravitational constant, 
$G_{\rm eff}(z,k)$ (see eq.~(\ref{cge})). 
The plots of $\delta_m(z)$ and $[f\sigma_8](z)$, shown in figures~\ref{deltaz} and~\ref{fsig8z}, 
were obtained by assuming $k=0.125$~Mpc$^{-1}$ and the Planck Collaboration best-fit $\sigma_{8,0}=0.811$~\cite{Planck1}.  

As in the background, increasing the model parameter $b$ causes the blue curves to overlap with the red ones, pointing out that this parameter provides a measurement of the similarity (or difference) between GR and $R^2$-AB model. 
Thus, the $f(R)$ model~(\ref{abcorrigido}) recovers GR whenever $R \gg R_{vac}$ or $b \gg 1$.


\section{\label{dataset}Cosmological datasets}

In this section, we briefly present the cosmological datasets used to constrain the free parameters of the $R^2$-AB model: $H(z)$ from Cosmic Chronometers (CC), $[f\sigma_8](z)$ from Red-shift-Space Distortions (RSD), and $m_B(z)$ from Pantheon+ and SH$0$ES.



\subsection{Cosmic Chronometers}

\begin{table}
\caption{The $31$ measurements on $H(z)$ obtained from the CC approach using the BC$03$ SPS model.}
    \centering
    \begin{tabular}{l l l}
    \hline\noalign{\smallskip}
        $z$       & $H(z)$           & Ref.                  \\
    \noalign{\smallskip}\hline\noalign{\smallskip}
        $0.07$    &  $69.0\pm 19.6$  &  \cite{Zhang2014}     \\   
        $0.09$    &  $69.0\pm 12.0$  &  \cite{Simon2005}     \\     
        $0.12$    &  $68.6\pm 26.2$  &  \cite{Zhang2014}     \\  
        $0.17$    &  $83.0\pm 8.0$   &  \cite{Simon2005}     \\  
        $0.179$   &  $75.0\pm 4.0$   &  \cite{Moresco2012}   \\   
        $0.199$   &  $75.0\pm 5.0$   &  \cite{Moresco2012}   \\  
        $0.2$     &  $72.9\pm 29.6$  &  \cite{Zhang2014}     \\   
        $0.27$    &  $77.0\pm 14.0$  &  \cite{Simon2005}     \\   
        $0.28$    &  $88.8\pm 36.6$  &  \cite{Zhang2014}     \\   
        $0.352$   &  $83.0\pm 14.0$  &  \cite{Moresco2012}   \\   
        $0.3802$  &  $83.0\pm 13.5$  &  \cite{Moresco2016}   \\   
        $0.4$     &  $95.0\pm 17.0$  &  \cite{Simon2005}     \\  
        $0.4004$  &  $77.0\pm 10.2$  &  \cite{Moresco2016}   \\    
        $0.4247$  &  $87.1\pm 11.2$  &  \cite{Moresco2016}   \\  
        $0.4497$  &  $92.8\pm 12.9$  &  \cite{Moresco2016}   \\  
        $0.47$    &  $89.0\pm 49.6$  &  \cite{Rats2017}      \\
        $0.4783$  &  $80.9\pm 9.0$    &  \cite{Moresco2016}  \\
        $0.48$    &  $97.0\pm 62.0$   &  \cite{Stern2010}    \\
        $0.593$   &  $104.0\pm 13.0$  &  \cite{Moresco2012}  \\ 
        $0.68$    &  $92.0\pm 8.0$    &  \cite{Moresco2012}  \\ 
        $0.781$   &  $105.0\pm 12.0$  &  \cite{Moresco2012}  \\
        $0.875$   &  $125.0\pm 17.0$  &  \cite{Moresco2012}  \\ 
        $0.88$    &  $90.0\pm 40.0$   &  \cite{Stern2010}    \\
        $0.9$     &  $117.0\pm 23.0$  &  \cite{Simon2005}    \\
        $1.037$   &  $154.0\pm 20.0$  &  \cite{Moresco2012}  \\
        $1.3$     &  $168.0\pm 17.0$  &  \cite{Simon2005}    \\
        $1.363$   &  $160.0\pm 33.6$  &  \cite{Moresco2015}  \\
        $1.43$    &  $177.0\pm 18.0$  &  \cite{Simon2005}    \\
        $1.53$    &  $140.0\pm 14.0$  &  \cite{Simon2005}    \\
        $1.75$    &  $202.0\pm 40.0$  &  \cite{Simon2005}    \\
        $1.965$   &  $186.5\pm 50.4$  &  \cite{Moresco2015}  \\
    \noalign{\smallskip}\hline
    \end{tabular}
    \label{hzdata}
\end{table}

One powerful technique to measure $H(z)$, independently of the assumption of a cosmological model, is the {\em cosmic
chronometers} method. 
This approach is based on the relationship
\begin{equation}
H\left(z\right) = -\frac{1}{\left(1+z\right)} \frac{dz}{dt} \simeq -\frac{1}{\left(1+z\right)} \frac{\Delta z}{\Delta t} \,,
\end{equation}
obtained from the definition $H \equiv \dot{a}/a$, where the derivative term, $dz / dt \simeq \Delta z / \Delta t$, can be determined from 
two passively-evolving galaxies, i.e., galaxies with old stellar populations and low star formation rates, whose {\em redshifts} are slightly 
different and whose ages are well-known. 
Furthermore, the chosen galaxies must have an age difference much smaller than their passively-evolving time~\cite{Jimenez2002}. 
Of course, to estimate the age of galaxies it is necessary to assume a {\em stellar population synthesis} (SPS) model. 
In table~\ref{hzdata} we list $31$ measurements on $H(z)$ obtained with the CC 
methodology, where the age of galaxies was obtained assuming BC$03$ SPS 
model~\cite{Bruzual2003}. 
Thus, these measurements contain systematic uncertainties related only to SPS model and to possible contamination due to the presence of young 
stars in quiescent galaxies~\cite{Gomez2019,Yang2020}.

\begin{table}
\caption{The $20$ measurements on $[f\sigma_8](z)$, obtained from several surveys and cosmological tracers, compiled by~\cite{Avila2022b}.}
    \centering
    \begin{tabular}{l l l}
    \hline\noalign{\smallskip}
        $z$      & $[f\sigma_8](z)$  & Ref.                \\
    \noalign{\smallskip}\hline\noalign{\smallskip}
         $0.02$  & $0.398 \pm 0.065$ & \cite{Turnbull2012} \\
         $0.025$ & $0.39 \pm 0.11$   & \cite{Achitouv2016} \\
         $0.067$ & $0.423 \pm 0.055$ & \cite{Beutler2012}  \\
         $0.10$  & $0.37 \pm 0.13$   & \cite{Feix2015}     \\
         $0.15$  & $0.53 \pm 0.16$   & \cite{Alam2017}     \\
         $0.32$  & $0.384 \pm 0.095$ & \cite{Sanchez}      \\
         $0.38$  & $0.497 \pm 0.045$ & \cite{Alam2017}     \\
         $0.44$  & $0.413 \pm 0.08$  & \cite{Blake2012}    \\
         $0.57$  & $0.453 \pm 0.022$ & \cite{Nadathur2019} \\
         $0.59$  & $0.488 \pm 0.060$ & \cite{Chuang2016}   \\
         $0.70$  & $0.473 \pm 0.041$ & \cite{Alam2017}     \\
         $0.73$  & $0.437 \pm 0.072$ & \cite{Blake2012}    \\
         $0.74$  & $0.50 \pm 0.11$   & \cite{Aubert2022}   \\
         $0.76$  & $0.440 \pm 0.040$ & \cite{Wilson2016}   \\
         $0.85$  & $0.52 \pm 0.10$   & \cite{Aubert2022}   \\
         $0.978$ & $0.379 \pm 0.176$ & \cite{Zhao2018}     \\
         $1.05$  & $0.280 \pm 0.080$ & \cite{Wilson2016}   \\
         $1.40$  & $0.482 \pm 0.116$ & \cite{Okumura2016}  \\
         $1.48$  & $0.30 \pm 0.13$   & \cite{Aubert2022}   \\
         $1.944$ & $0.364 \pm 0.106$ & \cite{Zhao2018}     \\
    \noalign{\smallskip}\hline
\end{tabular}
\label{fsig8zdata}
\end{table}

\subsection{Normalized growth rate}
The most common approach to study the clustering evolution of cosmic structures is through the normalized growth rate, $f\sigma_8$, given in eq.~(\ref{fsig8}). A possible way to obtain this one is by first measuring the velocity scale parameter, given by $v_\beta = f_g/\delta_{\text{bias}}$, where $\delta_{\text{bias}}$ is the {\em bias} factor, and then use it in the relation 
\begin{equation}
[f\sigma_8](z) = v_\beta(z)\,\sigma_8^{tr}(z)\, ,
\end{equation}
where $\sigma_8^{tr}(z)$ is the matter fluctuation amplitude of the cosmological tracer, e.g., HI line extra-galactic sources (EGS), luminous red galaxies (LRG), quasars (QSOs), type Ia Supernovae (SNe Ia), and emission-line galaxies (ELG).

The parametrized growth rate data, $[f\sigma_8](z)$, are most often obtained using the redshift-space distortions effect observed in galaxy surveys~\cite{Turnbull2012,Achitouv2016,Beutler2012,Feix2015,Alam2017,Sanchez,Avila2021,Avila2022a,Marques2020}. 
In table~\ref{fsig8zdata} we can see the data compilation from~\cite{Avila2022b} considering measurements of $[f\sigma_8](z)$. 
This compilation follows a methodology, in which double counting is eliminated and possible biases are reduced, thus ensuring the reliability of the dataset (see section $3$ of~\cite{Avila2022b}).

\subsection{Pantheon+ and SH0ES}

SNe Ia have been the most important tool in the exploration of the recent expansion history of the universe. 
The Supernovae have not only provided the initial confirmation of the accelerating expansion of the universe ~\cite{Riess1998,Perlmutter1999}, but today they also play a role in mapping the large-scale structure of the universe. 
With the growing abundance of SNe Ia observations at greater redshifts and advancements in analysis methods, cosmologists increasingly rely on them to explore the equation of state of dark energy.

From an observational point of view, it is assumed that different SNe Ia with identical color, shape (of the light curve), and galactic environment have on average the same intrinsic luminosity for all redshifts. 
This hypothesis is quantified through the empirical relationship~\cite{Brout2019,Trip1998}
\begin{equation}
    \mu_{obs} = m_B + \nu x_1 - \beta c + \delta_{\mu-\text{bias}} - M_B \,,
\end{equation}
where $\mu_{obs}$ is the observed distance modulus, $m_B$ correspond to observed peak magnitude in B-band rest-frame, while $\nu$, $\beta$, $\delta_{\mu-\text{bias}}$, and $M_B$ are the stretch of the light curve correction $x_1$, the SNe color at maximum brightness correction $c$, the simulated bias correction $\delta_{\mu-\text{bias}}$, and the absolute magnitude in the B-band rest-frame $M_B$, respectively~\cite{Trip1998,Brout2021,Popovic2021}.

On the other hand, the theoretical apparent magnitude $m_B$ for a bright source at redshift $z$ is given by
\begin{equation}
    m_B(z) = 5 \log{\left[\frac{d_L(z)}{10\, \text{pc}}\right]} + M_B \,,
\end{equation}
where $d_L(z)$ is the theoretical luminosity distance. The theoretical distance modulus reads as $\mu_{theo} = m_B - M_B$. For a flat cosmology ($K=0$), the luminosity distance is given by
\begin{equation}
d_L(z) = (1 + z)\int^z_0{\frac{d\Tilde{z}}{H(\Tilde{z})}}.
\end{equation}

\begin{figure}
\centering
\includegraphics[scale=0.35]{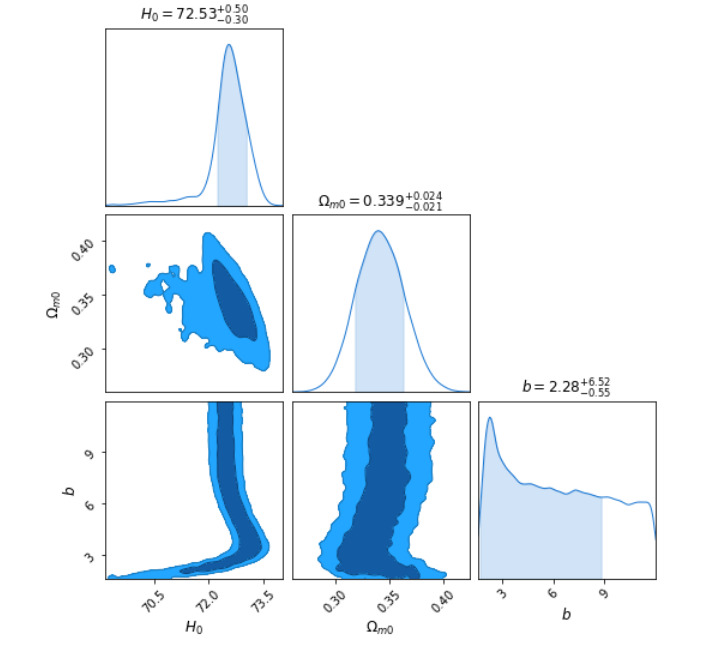}
\caption{MCMC analysis for the $R^2$-AB model considering only PN$^+$ SNe Ia data compilation. For this we have fixed the SHOES absolute magnitude, $M_B = -19.253$~\cite{riess2022comprehensive}, and the priors $H_0 = [64,\,76]$, $\Omega_{m,0}=[0.1,\,0.5]$, and $b=[1.6,\,12]$.} 
\label{snia_mcmc}
\end{figure}

In this regard, to test the $R^2$-AB model we have considerered the Pantheon$+$ (PN$^+$) compilation~\cite{Scolnic2022}, successor of the original Pantheon (PN)~\cite{Scolnic2018}, which have analysed $1701$ 
SNe Ia light curves with redshifts $0.001 \leq z \leq 2.26$. 
Due to the increase in sample size and better treatments of systematic uncertainties, the analysis with PN$^+$ presents an improvement factor of $2$ in the power of cosmological constraints in relation to the original PN~\cite{Scolnic2022}.


\section{\label{results}Analyses and Results}

\begin{figure*}
\centering
\includegraphics[scale=0.4]{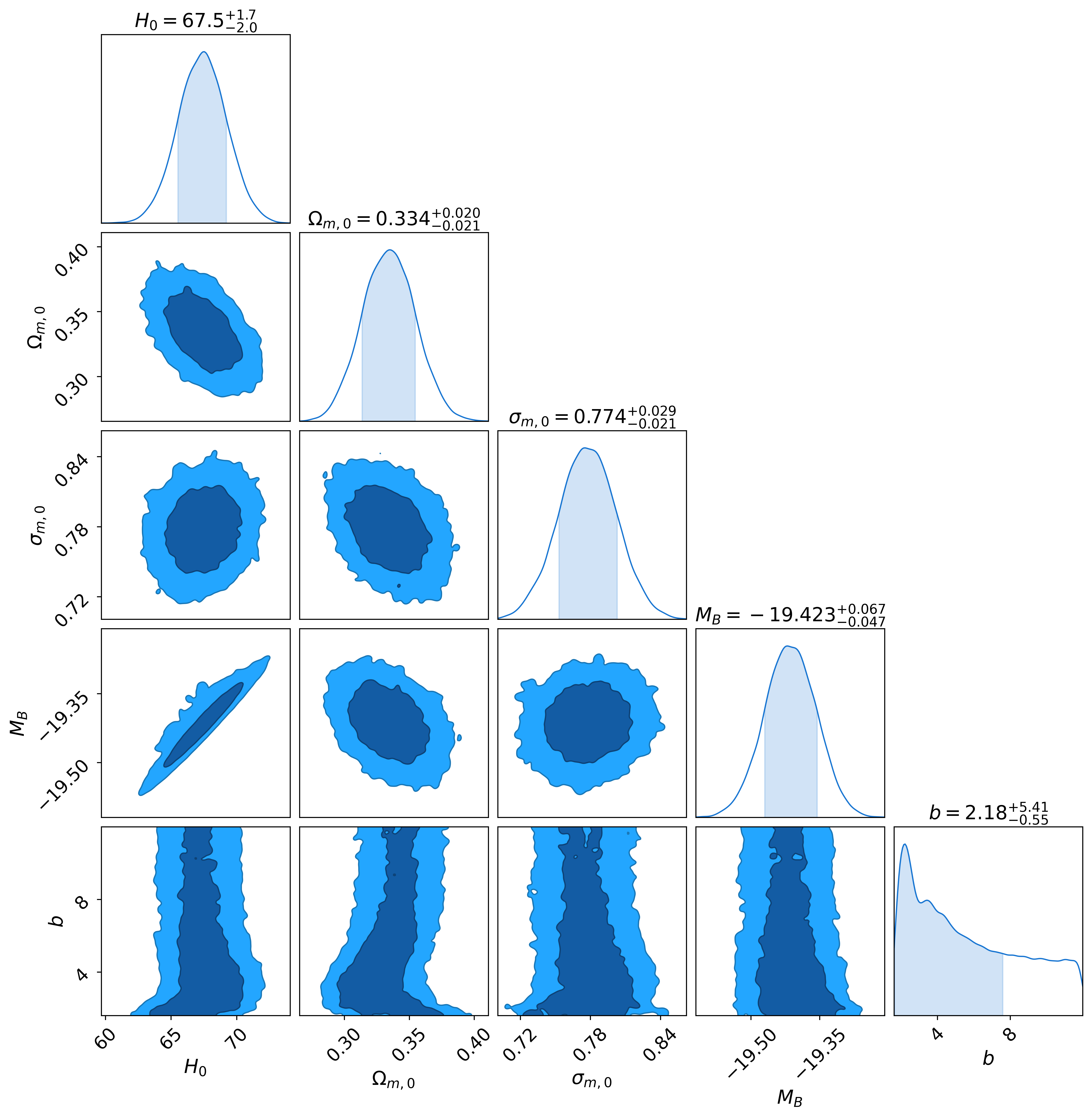}
\caption{MCMC analysis for the $R^2$-AB model considering the combination $\text{PN$^+$}+\text{CC}+\text{RSD}$. For this analysis we set the prior intervals $H_0 = [54,\,76]$, $\Omega_{m,0}=[0.1,\,0.5]$, $\sigma_{8,0}=[0.7,\,0.9]$,  $M_B=[-20.2,\,-19.0]$ and $b=[1.6,\,12]$.}
\label{mbzmcmc}
\end{figure*}



In our analyses, we take into account the Bayes' Theorem ~\cite{Robert2005}, which establishes a relationship between the probability of an event occurring and our prior knowledge of it. In simpler terms, it connects our knowledge of a specific parameter \emph{posteriori} (after obtaining data) with our \emph{a priori} knowledge (before observing the data):
\begin{equation}
P(\vartheta|D,\alpha) = \frac{P(D|\vartheta,\alpha)P(\vartheta|\alpha)}{P(D|\alpha)} \,,
\end{equation}
where $P(\vartheta|D,\alpha)$ is the {\em posterior} PDF, $P(D|\vartheta,\alpha)$ is the {\em likelihood}, $P(\vartheta|\alpha)$ is called {\em prior}, and $P(D|\alpha)$ is the {\em evidence}; moreover, $\vartheta$ represents the model parameters set, $D$ is the dataset and $\alpha$ denotes the {\em prior} information (model). Since the {\em evidence} $P(D|\alpha)$ is independent on model, we can ignore it as a normalizing constant. This approach offers a means to update our understanding of the parameter we aim to infer.

In order to generate random samples from a complex and high-dimensional probability distributions functions (PDF), we considered the \emph{Monte Carlo Markov Chain} (MCMC) technique, based in Metropolis-Hastings Algorithm. It starts with an initial sample and iteratively proposes new samples based on a proposal distribution. It then accepts or rejects the proposed sample based on an acceptance criterion that ensures the chain converges to the desired distribution. The {\em posterior} PDF getted around their most likely values allows us to obtain the best-fit model parameters with robust uncertainties.


If the observations $D_i$ are gaussian distributed the {\em likelihood} is given by the multivariate Gaussian~\cite{Verde2010},
\begin{equation}\label{like}
\mathcal{L}(\vartheta) = \exp{\left(-\frac{1}{2}\sum_{ij} \Delta E_{i}^{T}\, C^{-1}_{ij}\, \Delta E_{j}  \right)} \,,
\end{equation}
\begin{equation}
    \Delta E_{i} \equiv E_i(\vartheta|\alpha) - D_i\,,
\end{equation}
where $E_i(\vartheta|\alpha)$ is the ith expected value (based on a model) and $C_{ij}$ is the covariance matrix encoding statistical and systematic uncertainties related to dataset $D_i$. In case of strictly uncorrelated observations, this is simplified as $C_{ij}^{-1}=1/\sigma_i^2$, where $\sigma_i^2$ is the error at datum $i$.

The sum in eq.~(\ref{like}) is termed the {\em chi-square}, $\chi^2$, simplifying the form of the {\em likelihood} as $\mathcal{L} = \exp{(-\chi^2/2)}$. For a joint analysis of our datasets, the total chi-square is expressed by
\begin{equation}
    \chi^2 = \chi_\text{SNe}^2 + \chi_\text{CC}^2 + \chi_\text{RSD}^2\,,
\end{equation}
resulting in the total {\em likelihood} $\mathcal{L} = \mathcal{L}_\text{SNe} \times \mathcal{L}_\text{CC} \times \mathcal{L}_\text{RSD}$.

Additionally, it is usual to consider that the {\em prior} sets have the same probability of occurrence, so that $P(\vartheta|\alpha)$ has the form of the Dirac delta distribution,
\begin{equation}
P(\vartheta|\alpha) = \left\{
\begin{array}{ll}
    1 & \text{if} \quad \vartheta^{(0)} < \vartheta < \vartheta^{(1)}, \\
    0 & \text{otherwise},
    \end{array}
    \right.
\end{equation}
where $\vartheta^{(0)}$ and $\vartheta^{(1)}$ are the flat {\em prior} intervals. 
Hence the {\em posterior} PDF is given by  $P(\vartheta|D,\alpha) = \mathcal{L}(\vartheta)$, i.e, only by the {\em likelihood}.

There are two ways to estimate the optimum value for the model parameters: maximum {\em likelihood} and least {\em chi-square}. 
The latter is more common and can be done using a Python library (e.g., {\em scipy.optimize}). 
Next, we shall explore the parametric space of the model parameters, sampling the {\em posterior} distribution around that value, following the MCMC method and the Metropolis-Hastings algorithm. 
The confidence regions has been drawn assuming $\mathcal{L} = \mathcal{L}_{max} + {\Delta\mathcal{L}}_0$, where the constant ${\Delta\mathcal{L}}_0$ is determined by the cumulative probability density. To implement the MCMC routine we use Python as well.

\begin{table}
\caption{Best-fit values from our MCMC likelihood analyses.}
\centering
\begin{tabular}{c c c}
\hline\noalign{\smallskip}
Parameter & Only SNe & SNe+CC+RSD \\
\noalign{\smallskip}\hline\noalign{\smallskip}
$H_0$          & $72.53^{+0.5}_{-0.3}$     & $67.5^{+1.7}_{-2.0}$        \\
$\Omega_{m,0}$ & $0.339^{+0.024}_{-0.021}$ & $0.334^{+0.020}_{-0.021}$   \\
$\sigma_{8,0}$ & ---                       & $0.774^{+0.029}_{-0.021}$   \\
$M_B$          & fixed                     & $-19.423^{+0.067}_{-0.047}$ \\
$b$            & $2.28^{+6.52}_{-0.55}$    & $2.18^{+5.41}_{-0.55}$      \\
\noalign{\smallskip}\hline
\end{tabular}
\label{table_results}
\end{table}

Note that $[f\sigma_8]$ and $G_{\rm eff} / G$ are independent of $H_0$. 
In our analyses we set: 
\begin{itemize}
\item
$\vartheta=[H_0,\,\Omega_{m,0},\,b]$, for the Hubble function;
\item
$\vartheta=[\Omega_{m,0},\,\sigma_{8,0},\,b]$, for the growth rate; 
and 
\item
$\vartheta=[H_0,\, \Omega_{m,0},\, M_B,\, b]$, for the apparent magnitude.
\end{itemize}
The priors were defined according to the analysis carried out.

In figures~\ref{snia_mcmc} and~\ref{mbzmcmc} we show the results of our MCMC analyses considering only PN$^+$ SNe Ia dataset and the combination $\text{SNe}+\text{CC}+\text{RSD}$, respectively, for the $R^2$-AB model; we summarize these results in table~\ref{table_results}. 
As seen in this table, the most likely values obtained for the model parameters depends on the analysis. 
Because of the $H_0-M_B$ degeneracy, for only SNe Ia data we set $M_B=-19.253$, which is the value of absolute magnitude measured by SH$0$ES collaboration~\cite{riess2022comprehensive}, compatible with local universe, resulting in $H_0 = 72.53^{+0.5}_{-0.3}~\text{km s}^{-1}\text{Mpc}^{-1}$, $\Omega_{m,0}=0.339^{+0.024}_{-0.021}$, and $b=2.28^{+6.52}_{-0.55}$. 
Instead, in the joint analysis we can see that this degeneracy is broken by CC $H(z)$ data, then $H_0=67.5^{+1.7}_{-2.0}~\text{km s}^{-1}\text{Mpc}^{-1}$, $\Omega_{m,0}=0.334^{+0.020}_{-0.021}$, $\sigma_{8,0}=0.774^{+0.029}_{-0.021}$, $b=2.18^{+5.41}_{-0.55}$, and the absolute magnitude obtained is $M_B=-19.423^{+0.067}_{-0.047}$.

Although the $\chi^2$ statistics is effective in searching for the {\em best-fit} parameters whithin a given model, it is not suitable for setting up comparisons between models with different number of parameters because lower $\chi^2$ values can be obtained increasing the number of parameters. 
Accordingly, other criteria for model selection are used in the literature, such as {\em Akaike Information Criterion} (AIC)~\cite{Akaike1974} (see also~\cite{Motohashi2013} for an application of the AIC criterion to $f(R)$ cosmology) and 
$\Bar{\chi}^2_\text{min} \equiv \chi^2_\text{min}/(N-\Sigma)$, where $N$ and $\Sigma$ are the number of data points and the number of independently adjusted parameters, 
respectively.

An information criterion AIC of $\vartheta$ is defined as
\begin{equation}
    AIC(\vartheta) \equiv \chi^2_\text{min}(\vartheta) + 2\Sigma\,,
\end{equation}
where $\chi^2_\text{min} \equiv -2 \ln{(\mathcal{L}_{max})}$. 
Thus, the difference between the investigated model and a referring model (naturally, the $\Lambda$CDM) can be measured as $\Delta AIC = \Delta\chi_\text{min}^2 + 2\Delta\Sigma$.

In order to make a comparison, we consider the $\Lambda$CDM model as a referring model, for this we have analysed it with the same observational data as the 
$R^2$-AB model. 
Our resuls are summarized in table~\ref{table_results2}. 
One observes that the AIC criterion ends up penalizing the $R^2$-AB model since it has $1$ more independent parameter, therefore according to this criterion the $\Lambda$CDM is the model that best-fits the cosmological data analysed, i.e., 
the datasets of SNe$+$CC$+$RSD. 

\begin{table}
\caption{$\chi^2$ statistics for model comparison: $\Lambda$CDM {\em versus} $R^2$-AB model. 
As observed, the AIC criterion penalizes the $R^2$-AB 
model because it has 1 more independent parameter, therefore 
the $\Lambda$CDM is the model that best-fits the cosmological data analysed.
}
\centering
\begin{tabular}{l l l}
\hline\noalign{\smallskip}
                                & \multicolumn{2}{c}{Models}    \\
\cmidrule{2-3}
\multirow[c]{-2}{*}{Estimators} & $\Lambda$CDM   & $R^2$-AB     \\
\noalign{\smallskip}\hline\noalign{\smallskip}
$\chi^2_\text{min}$             & $2898.865$     & $2897.581$   \\
$N-\Sigma$                      & $1748$         & $1747$       \\
$\Bar{\chi}^2_\text{min}$       & $1.658$        & $1.659$      \\
$\Delta AIC$                    & $0$            & $0.716$      \\
\noalign{\smallskip}\hline
\end{tabular}
\label{table_results2}
\end{table}



\section{\label{conclusions}Conclusions}

The flat-$\Lambda$CDM, with its mysterious dark energy component in the form of a cosmological constant, is not the final model of cosmology. 
Efforts are being devoted to study alternative cosmological scenarios where GR theory is modified in a way to explain the observed universe, both at the background and perturbative levels, but having GR as a suitable limit. 

A class of possible candidates to explain the accelerated expansion of the universe is based on interpretation of gravity slightly different from that provided by GR, a theoretical approach known by the generic name of {\it modified gravity theory} (MG theory). 
This new geometric scenario for the space-time has to satisfy phenomenological 
rigorous criteria~\cite{Capozziello2011,Clifton2012,Faraoni2010,Papantonopoulos2014}: 
\begin{itemize}
\item because GR is a well-established theory for the strong gravitational fields and small scales, any attempt to modify it should contain GR as a limiting theory at suitable scales and strong gravitational fields; 
\item in the distant past, $z \gg 1$, the MG theory should have a behavior concordant with a matter dominated era;
\item at large scales and from a recent past, $z < z_t$ (= transition redshift), the behavior expected is such that the MG theory explains the accelerated expansion phase of the universe, a feature well established by different 
cosmological tracers (background level);
\item at the perturbative level, the MG theory should satisfactorily explain the growth 
rate of cosmic structures data. 
\end{itemize}
This cosmic phenomenology, expected to be satisfied by any contender of the concordance model of cosmology, the $\Lambda$CDM model, makes non-trivial the search for good $f(R)$ candidates. 
One such viable candidate is the $R^2$-AB model, here investigated using cosmological data to constrain its model parameters~\cite{Appleby2007,Appleby2010,Appleby2008}. 
A general criticism concerning $f(R)$ models is their tendency to exhibit unbounded growth of the scalaron mass at high energies, or in the early universe, introducing instabilities in the model. 
This issue can be addressed through the addition of an $R^2$ term, which effectively constrains the scalaron mass preventing an excessive growth, ensuring the viability of the 
model~\cite{DeFelice2010,Starobinsky2007,Appleby2010}. 


In general, a $f(R)$ or other alternative model undergo the problem that having a large number of parameters makes the statistical best-fit process 
be less efficient than that one made with the flat-$\Lambda$CDM with just one free parameter. 
For this our interest here in studying the corrected Appleby-Battye model, the 
$R^2$-AB model, a model with 2 parameters and 1 constraining relation, 
which determines a model with one free parameter: $b$. 
In this way, we performed MCMC analyses of the $R^2$-AB model and observed that, 
for both $H(z)$ and $[f\sigma_8](z)$ studies, the values obtained for $H_0$, $\Omega_{m,0}$, and $\sigma_{8,0}$ were fully concordant with the flat-$\Lambda$CDM cosmological parameters obtained 
by the Planck Collaboration~\cite{Planck1}. 


Moreover, our MCMC statistical analyses of the section~\ref{results} show that the $R^2$-AB model is reasonably well constrained by the cosmological data applied: (i) using only PN$^+$ SNe Ia data the analysis returns a best-fit value for the model parameter $b = 2.28^{+6.52}_{-0.55}$, and (ii) the joint analysis SNe+CC+RDS returns $b = 2.18^{+5.41}_{-0.55}$, with both values compatible with each other and within the interval where the $R^2$-AB model satisfies all the phenomenological criteria mentioned above~\cite{Amendola2006,Appleby2007,Appleby2010}.


The result of our analyses shows that the $R^2$-AB model is consistent with observational data, including both the background and perturbative aspects. However, the determination of model parameter was inconclusive (see table~\ref{table_results}). 
This situation highlights the need for further investigation into alternative scenarios and additional analyses incorporating different observational datasets.


\begin{figure}
\centering
\includegraphics[scale=0.445]{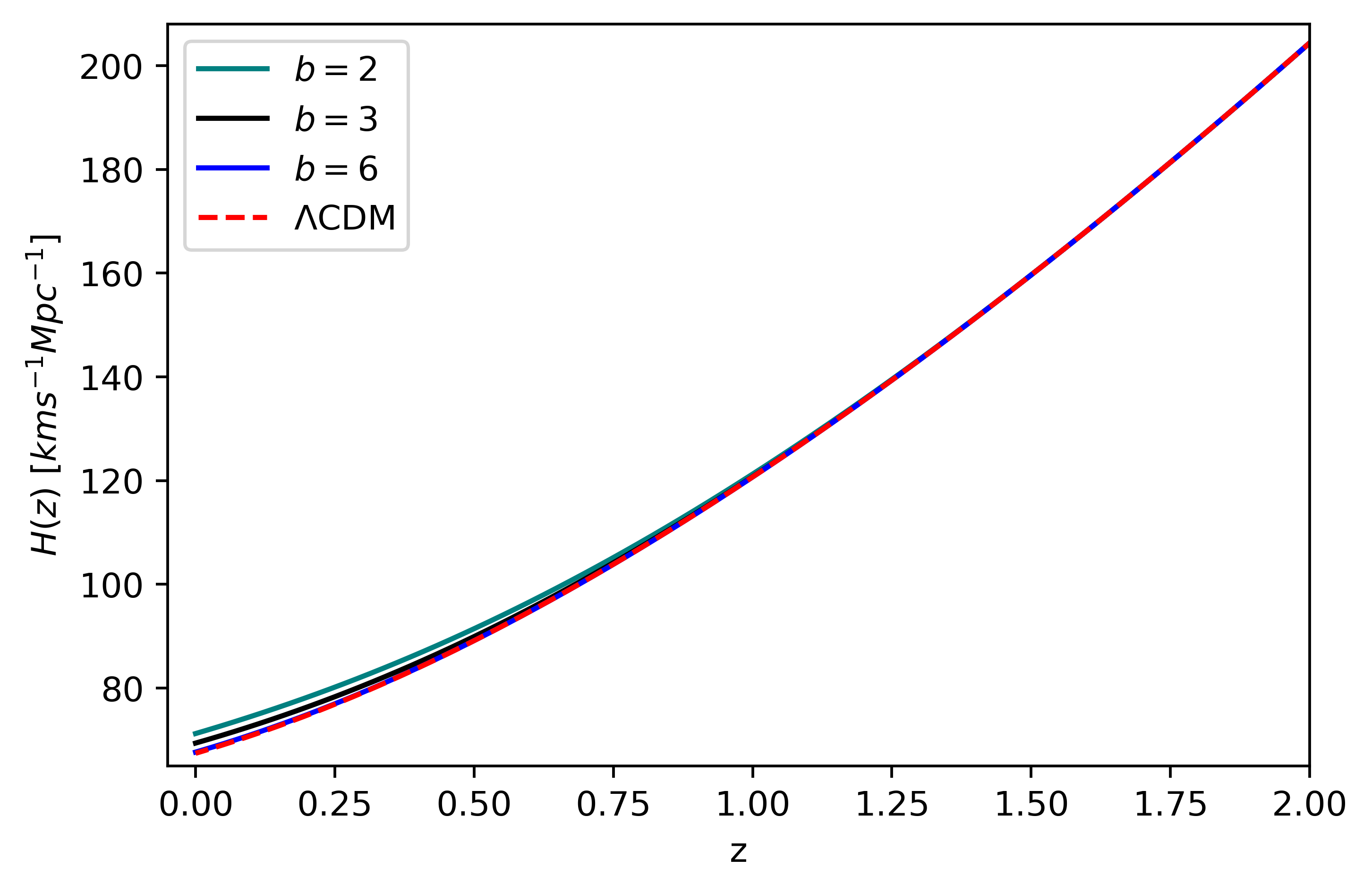}
\caption{Hubble function $H(z)$ for different values of parameter $b$ and setting $H_0=67.4$ km/s/Mpc and $\Omega_{m,0}=0.315$. The dashed red curve corresponds to flat-$\Lambda$CDM model.}
    \label{hb}
\end{figure}
\begin{figure}
    \centering
    \includegraphics[scale=0.445]{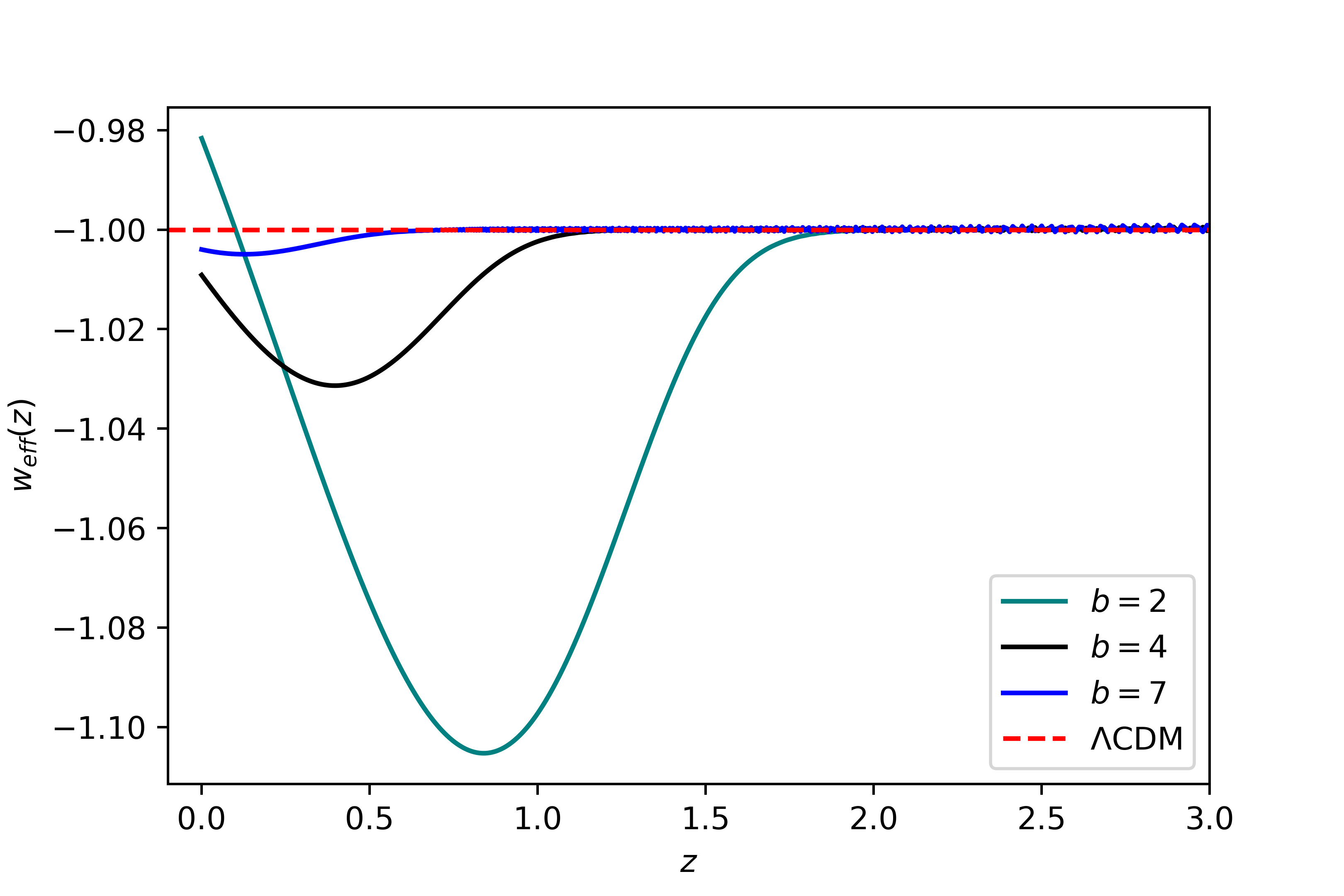}
\caption{Equation of state for different 
values of parameter $b$ and setting $H_0=67.4$ km/s/Mpc 
and $\Omega_{m,0}=0.315$. The dashed red line corresponds to 
$w_{\Lambda} = -1$ from flat-$\Lambda$CDM model.}
\label{wb}
\end{figure}
\begin{figure}
\centering
\includegraphics[scale=0.445]{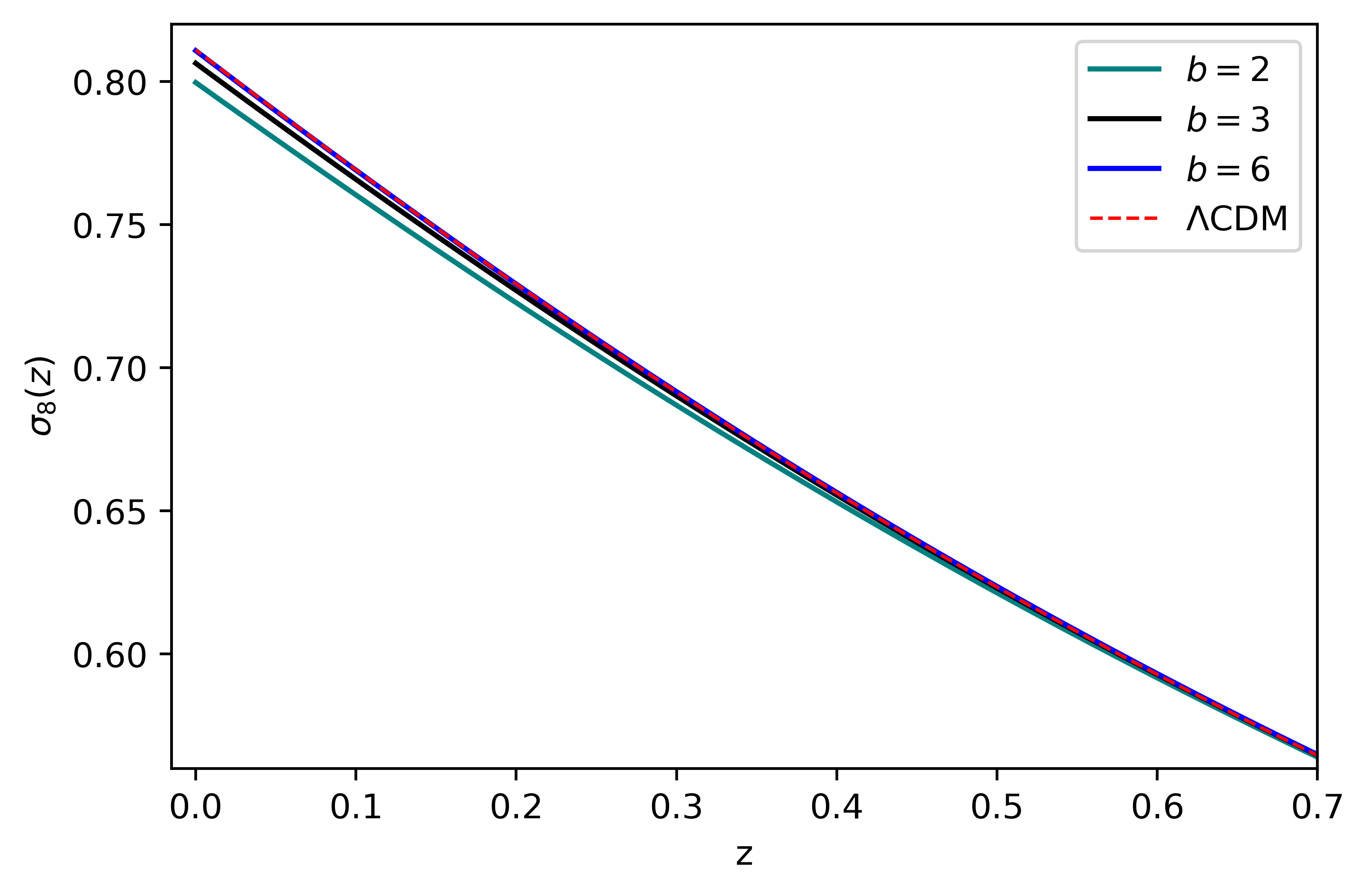}
\caption{$\sigma_8(z)$ function considering diferent values of parameter $b$ and setting $\Omega_{m,0}=0.315$ and $\sigma_{8,0}=0.811$. The dashed red curve corresponds to flat-$\Lambda$CDM model.}
    \label{sig8b}
\end{figure}
\begin{figure}
\centering
\includegraphics[scale=0.445]{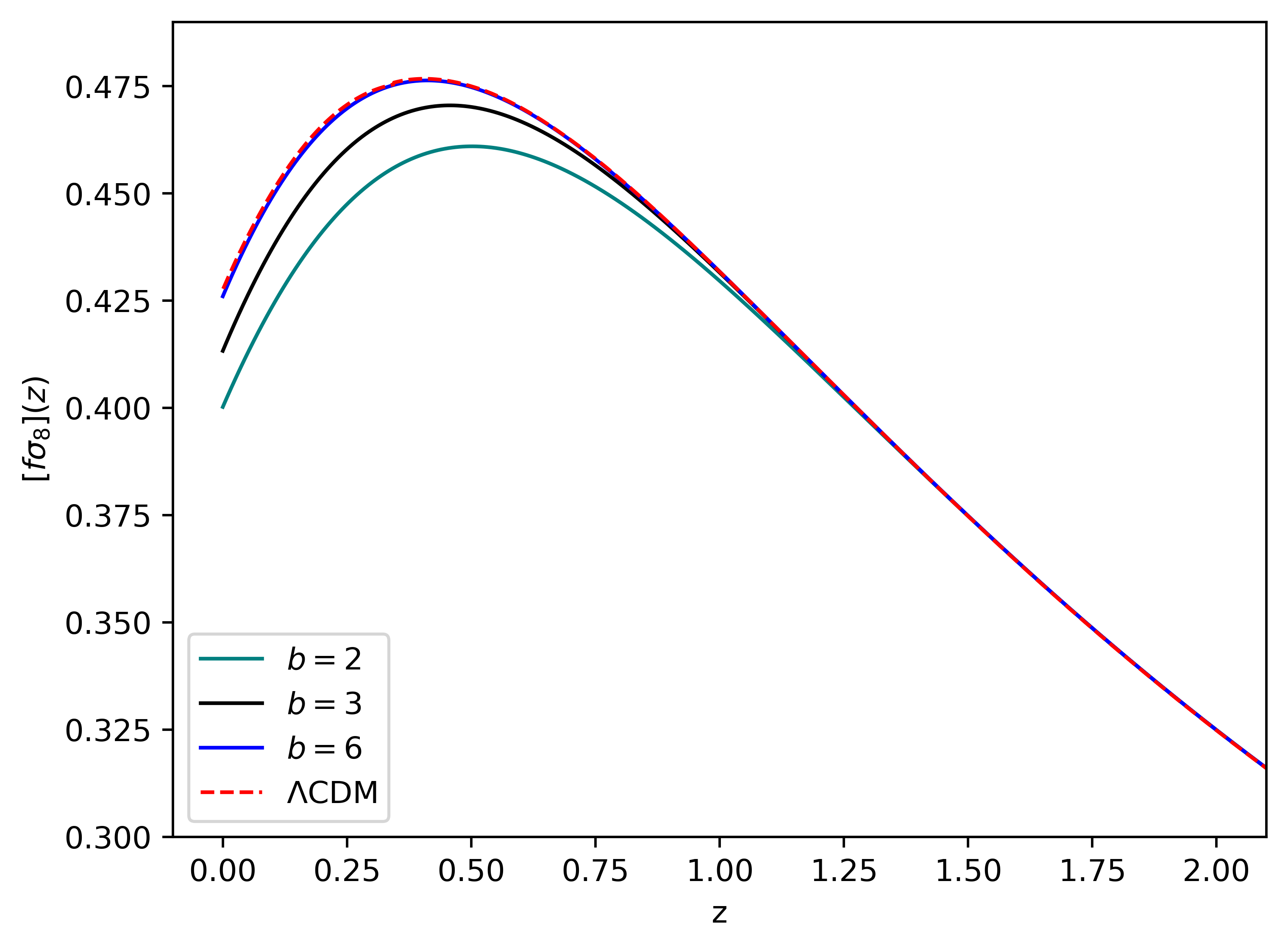}
\caption{$[f\sigma_8](z)$ for diferent values of parameter b and setting
$\Omega_{m,0}=0.315$ and $\sigma_{8,0}=0.811$. The dashed red curve corresponds to flat-$\Lambda$CDM case.}
    \label{fsig8b}
\end{figure}


\section*{Acknowledgments}

BR and AB would like to thank the Brazilian Agencies CAPES and CNPq for their respective fellowships. MC would like to acknowledges the Observat\'orio Nacional for the hospitality.


\appendix
\section{\label{apen} Effect of different $b$ values on cosmic observables}

We find interesting to compare the evolution of some observables, like 
$H,\, w_{\rm eff},\, \sigma_8$, and $[f \sigma_8]$, for different values of the 
model parameter $b$. 
For this, in figures~\ref{hb}, \ref{wb}, \ref{sig8b} and~\ref{fsig8b} 
we show these plots to observe the effect of $b$ on some cosmological 
observables obtained in the $R^2$-AB model. 
These figures illustrate this comparative analysis, where we are keeping the 
values of the other parameters, $H_0$, $\Omega_{m,0}$, $\sigma_{8,0}$, fixed; 
moreover, we keep the scale mass as $\Delta=10^{-7}$.

\section{\label{apen2}
Comparison of R$^2$-AB with Hu-Sawicki and Starobinsky models}

The Hu-Sawicki~\cite{Hu2007} and Starobinsky~\cite{Starobinsky2007} models can be expressed as
\begin{equation}
    f_{HS}(R) = R - R_{HS}\frac{c_1(R/R_{HS})^n}{c_2(R/R_{HS})^n + 1} \,,
\end{equation}
\begin{equation}
    f_{S}(R) = R + \lambda R_s \left[\left(1 + \frac{R^2}{R_s^2}\right)^{-n} - 1\right] \,,
\end{equation}  
where $R_{HS} \equiv H_0^2\Omega_{m,0}$ and $R_s$ correspond to the present curvature scale for each of the models, respectively.
%
%
The $\Lambda\text{CDM}$ limit, i.e., $R \gg R_0$, implies in the constraints 
\begin{equation}
c_1 = 6c_2\frac{1-\Omega_{m,0}}{\Omega_{m,0}}\,, \quad  
R_s = \frac{6H_0^2(1-\Omega_{m,0})}{\lambda} \,,    
\end{equation}
relating $c_1$ to $c_2$ and $R_s$ to $\lambda$. This means that the real free parameters of these models are 2: $\{n,c_2\}$ for the Hu-Sawicki and $\{n,\lambda\}$ for the Starobinsky models, whereas the $R^2$-AB model contains only $1$, that is, $\{b\}$. 


In addition to the number of free parameters, another difference from the Hu-Sawicki and Starobinsky models to the Appleby-Battye model is that: both Hu-Sawicki and Starobinsky models contain power law corrections to GR, whereas Appleby-Battye model contains exponentially suppressed corrections~\cite{Hu2007,Starobinsky2007,Appleby2007}. 
Moreover, because $f_{AB}^{\prime\prime}$, in non-corrected AB model, vanishes much more rapidly than $f_{HS}^{\prime\prime}$ or $f_{S}^{\prime\prime}$, we expect that the singularity ($\mathcal{M}_s \rightarrow \infty$) forms earlier in AB model.


\begin{thebibliography}{}
%
%

\bibitem{Riess1998}
A.~G.~Riess et al., AJ \textbf{116}, (1998) 1009.

\bibitem{Perlmutter1999}
S.~Perlmutter et al., AJ \textbf{517}, (1999) 565.

\bibitem{Riess2022}
A.~G.~Riess et al., AJ \textbf{938}, (2022) 36.

\bibitem{Anchordoqui2021}
L.~A.~Anchordoqui, E.~Di~Valentino, S.~Pan and W.~Yang, JHEAP \textbf{32}, (2021) 28.

\bibitem{Yang:2021eud}
W.~Yang, E.~Di~Valentino, S.~Pan, A.~Shafieloo and X.~Li, Phys. Rev. D \textbf{104}, (2021) 063521.

\bibitem{Odintsov:2020qzd}
S.~D.~Odintsov, D.~Sáez-Chillón Gómez and G.~S.~Sharov, Nucl. Phys. B \textbf{966}, (2021) 115377.

\bibitem{Bernui2023}
A.~Bernui, E.~Di~Valentino, W.~Giar{\`e}, S.~Kumar and R.~C.~Nunes, Phys. Rev. D \textbf{107}, (2023) 103531.

\bibitem{Capozziello2011}
S.~Capozziello and M.~De~Laurentis, Phys. Rep. \textbf{509}, (2011) 167.

\bibitem{Clifton2012}
T. Clifton and P.~G.~Ferreira, A.~Padilla and C.~Skordis, Phys. Rep. \textbf{513}, (2012) 1.

\bibitem{DeFelice2010}
A.~De~Felice and S.~Tsujikawa, Living Rev. Relativ. \textbf{13}, (2010) 1.

\bibitem{Sotiriou2008}
T.~P.~Sotiriou and V.~Faraoni, RMP \textbf{82}, (2010) 451.

\bibitem{Starobinsky1980}
A.~Starobinsky, Phys. Lett. B \textbf{91}, (1980) 99.

\bibitem{Amendola2006}
L.~Amendola, R.~Gannouji, D.~Polarski and S.~Tsujikawa, Phys. Rev. D \textbf{75}, (2007) 083504.

\bibitem{Hu2007}
W.~Hu and I.~Sawicki, Phys. Rev. D \textbf{76}, (2007) 064004.

\bibitem{Starobinsky2007}
A.~Starobinsky, JETP Lett. \textbf{86}, (2007) 157.

\bibitem{Appleby2007}
S.~Appleby and R.~Battye, Phys. Lett. B \textbf{654}, (2007) 7.

\bibitem{Li2007}
B.~Li and J.~Barrow, Phys. Rev. D \textbf{75}, (2007) 084010.

\bibitem{Amendola2008}
L.~Amendola and S.~Tsujikawa, Phys. Lett. B \textbf{660}, (2008) 125.

\bibitem{Tsujikawa2008}
S.~Tsujikawa, Phys. Rev. D \textbf{77}, (2008) 023507.

\bibitem{Cognola2008}
G.~Cognola et al., Phys. Rev. D \textbf{77}, (2008) 046009.

\bibitem{Linder2009}
E.~V.~Linder, Phys. Rev. D \textbf{80}, (2009) 123528.

\bibitem{Elizalde2011}
E.~Elizalde, S.~Nojiri, S.~D.~Odintsov, L.~Sebastiani and S.~Zerbini, Phys. Rev. D \textbf{83}, (2011) 086006.

\bibitem{Xu2014}
Q.~Xu and B.~Chen, Commun. Theor. Phys. \textbf{61}, (2014) 141.

\bibitem{Nautiyal2018}
A.~Nautiyal, S.~Panda and A.~Patel, Int. J. Mod. Phys. D \textbf{27}, (2018) 1750185.

\bibitem{Gogoi2020}
D.~Gogoi and U.~Goswami, Eur. Phys. J. C \textbf{80}, (2020) 1.

\bibitem{Oikonomou2013}
V.~K.~Oikonomou, Gen. Relativ. Gravit. \textbf{45}, (2013) 2467.

\bibitem{Oikonomou2021}
V.~K.~Oikonomou, Phys. Rev. D \textbf{103}, (2021) 044036.


\bibitem{Appleby2010}
S.~Appleby, R.~Battye and A.~Starobinsky, JCAP \textbf{06}, (2010) 005.




\bibitem{Avila2019}
F.~Avila, C.~Novaes, A. Bernui, E.~de~Carvalho and J.~P.~Nogueira-Cavalcante, MNRAS \textbf{488}, (2019) 1481.

\bibitem{Marques2020}
G.~A.~Marques and A.~Bernui, JCAP \textbf{05}, (2020) 052.

\bibitem{deCarvalho2021}
E.~de~Carvalho, A.~Bernui, F.~Avila, C.~P.~Novaes and J.~P.~Nogueira-Cavalcante, A\&A \textbf{649}, (2021) A20.

\bibitem{Franco23}
C. Franco, F. Avila and A. Bernui, MNRAS \textbf{527}, (2024) 7400.

\bibitem{Oliveira23}
F. Oliveira, F. Avila, A. Bernui, A. Bonilla and R. C. Nunes, EPJC \textbf{84}, (2024) 636.

\bibitem{Appleby2008}
S.~Appleby and R.~Battye, JCAP \textbf{05}, (2008) 019.

\bibitem{Faraoni2010}
V.~Faraoni and S.~Capozziello, \textit{Beyond Einstein Gravity: A Survey of Gravitational Theories for Cosmology and Astrophysics} (Springer, Dordrecht 2011) 428.

\bibitem{Papantonopoulos2014}
E.~Papantonopoulos, \textit{Modifications of Einstein's Theory of Gravity at Large Distances}, (Springer, Switzerland 2014) 442.

\bibitem{Muller1987}
V.~Muller, H.~J.~Schmidt and A.~Starobinsky, Phys. Lett. B \textbf{202}, (1988) 198.

\bibitem{Nunez2004}
A.~Nunez and S.~Solganik, arXiv:hep-th/0403159.

\bibitem{Krause2006}
A.~Krause and S.~Ng, Int. J. Mod. Phys. A \textbf{21}, (2006) 1091.

\bibitem{Himmetoglu2009}
B.~Himmetoglu, C.~R.~Contaldi and M.~Peloso, Phys. Rev. D \textbf{80}, (2009) 123530.

\bibitem{Deruelle2011}
N.~Deruelle, M.~Sasaki, Y.~Sendouda and A.~Youssef, JCAP \textbf{2011}, (2011) 040.

\bibitem{Dolgov2003}
A.~Dolgov and M.~Kawasaki,
Phys. Lett. B \textbf{573}, (2003) 1.

\bibitem{Olmo2005}
G.~J.~Olmo, Phys. Rev. D \textbf{72}, (2005) 083505.

\bibitem{Faraoni2006}
V.~Faraoni, Phys. Rev. D \textbf{74}, (2006) 023529.



\bibitem{Hu2002}
W.~Hu and S.~Dodelson, Annu. Rev. Astron. Astrophys. \textbf{40}, (2002) 171.

\bibitem{Sasaki1984}
H.~Kodama and M.~Sasaki, Prog. Theor. Phys. Supplement \textbf{78}, (1984) 1.

\bibitem{Mukhanov1992}
V.~F.~Mukhanov, H.~A.~Feldman and R.~H.~Brandenberger, Phys. Rep. \textbf{215}, (1992) 203.

\bibitem{Bardeen1882}
J.~M.~Bardeen, Phys. Rev. D \textbf{22}, (1980) 1882.

\bibitem{peebles1967}
P.~J.~E.~Peebles, ApJ \textbf{147}, (1967) 859.

\bibitem{zel1970}
Y.~B.~Zel'Dovich, A$\&$A \textbf{5}, (1970) 84.

\bibitem{Tsujikawa2007}
S.~Tsujikawa, Phys. Rev. D \textbf{76}, (2007) 023514.

\bibitem{Strauss1995}
M.~Strauss and J.~Willick, Phys. Rep. \textbf{261}, (1995) 271.

\bibitem{Wang1998}
L.~Wang and P.~J.~Steinhardt, ApJ \textbf{508}, (1998) 483.

\bibitem{Linder2007}
E.~Linder and R.~Cahn, Astropart. Phys. {\bf 28}, (2007) 4.

\bibitem{Nesseris2017}
S.~Nesseris, G.~Pantazis and L.~Perivolaropoulos, Phys. Rev. D \textbf{96}, (2017) 023542.

\bibitem{Motohashi2012}
H.~Motohashi and A.~ Nishizawa, Phys. Rev. D \textbf{86}, (2012) 083514.

\bibitem{Motohashi2014}
A.~ Nishizawa and H.~Motohashi, Phys. Rev. D \textbf{89}, (2014) 063541.

\bibitem{Motohashi2010}
H.~Motohashi and A.~Starobinsky and J.~Yokoyama, Prog. Theor. Phys. \textbf{123}, (2010) 887.

\bibitem{Motohashi2013}
H.~Motohashi and A.~Starobinsky and J.~Yokoyama, Phys. Rev. Lett. \textbf{110}, (2013) 121302.

\bibitem{Planck1}
N.~Aghanim el al., A$\&$A \textbf{641}, (2020) A6.

\bibitem{Jimenez2002}
R.~Jimenez and A.~Loeb, ApJ \textbf{573}, (2002) 37.

\bibitem{Bruzual2003}
G.~Bruzual and S.~Charlot, MNRAS \textbf{344}, (2003) 1000.

\bibitem{Gomez2019}
A.~Gómez-Valent, JCAP \textbf{05}, (2019) 026.

\bibitem{Yang2020}
Y.~Yang and Y.~Gong, JCAP \textbf{2020}, (2020) 059.

\bibitem{Turnbull2012}
S.~J.~Turnbull et al., MNRAS \textbf{420}, (2012) 447.

\bibitem{Achitouv2016}
I.~Achitouv, C.~Blake, P.~Carter, J.~Koda and F.~Beutler, Phys. Rev. D \textbf{95} (2016) 083502.

\bibitem{Beutler2012}
F.~Beutler et al., MNRAS \textbf{423}, (2012) 3430.

\bibitem{Feix2015}
M.~Feix, A.~Nusser and E.~Branchini, Phys. Rev. Lett. \textbf{115}, (2015) 011301.

\bibitem{Alam2017}
S.~Alam et al., MNRAS \textbf{470}, (2017) 2617.

\bibitem{Sanchez}
A.~G.~Sánchez et al., MNRAS \textbf{440}, (2014) 2692.

\bibitem{Avila2021}
F.~Avila, A.~Bernui, E.~de~Carvalho and C.~P.~Novaes, MNRAS \textbf{505}, (2021) 3404.

\bibitem{Avila2022a}
F.~Avila, A.~Bernui, R.~C.~Nunes, E.~de~Carvalho and C.~P.~Novaes, MNRAS \textbf{509}, (2022) 2994.

\bibitem{Zhang2014}
C.~Zhang et al, Res. Astron. Astrophys. \textbf{14}, (2014) 1221.

\bibitem{Simon2005}
J.~Simon, L.~Verde and R.~Jimenez, Phys. Rev. D \textbf{71}, (2005) 123001.

\bibitem{Moresco2012}
M.~Moresco et al., JCAP \textbf{08}, (2012) 006.

\bibitem{Moresco2016}
M.~Moresco et al., JCAP \textbf{05}, (2016) 014.

\bibitem{Rats2017}
A.~Ratsimbazafy et al., MNRAS \textbf{467}, (2017) 3239.

\bibitem{Stern2010}
D. Stern, R.~Jimenez, L.~Verde, M.~Kamionkowski and S.~A.~Stanford, JCAP \textbf{2010}, (2010) 008.

\bibitem{Moresco2015}
M.~Moresco, MNRAS Lett. \textbf{450}, (2015) L16.

\bibitem{Avila2022b}
F.~Avila, A.~Bernui, A.~Bonilla and R.~C.~Nunes, EPJC \textbf{82}, (2022) 594.

\bibitem{Brout2019} 
D.~Brout et al., AJ \textbf{874}, (2019) 150.

\bibitem{Trip1998} 
R.~Tripp, A\&A \textbf{331}, (1998) 815.

\bibitem{Brout2021} 
D.~Brout and D. Scolnic, AJ \textbf{909}, (2021) 26.

\bibitem{Popovic2021}
B.~Popovic, D.~Brout, R. Kessler, D. Scolnic and L. Lu, AJ \textbf{913}, (2021) 49.

\bibitem{Scolnic2022}
D. Scolnic et al., AJ \textbf{938}, (2022) 113.

\bibitem{Scolnic2018}
D. Scolnic et al., AJ \textbf{859}, (2018) 101.

\bibitem{riess2022comprehensive}
A.~Riess, et al., AJ lett. \textbf{934}, (2022) L7.

\bibitem{Robert2005}
C.~Robert and G.~Casella, \textit{Monte Carlo Statistical Methods} (Springer, New York, 2005) 647.

\bibitem{Verde2010}
L.~Verde, \textit{Statistical methods in cosmology} (Springer, Berlin Heidelberg, 2010) 30.

\bibitem{Akaike1974}
H.~Akaike, IEEE Trans. Automat. Contr. \textbf{19}, (1974) 716.

\bibitem{Blake2012}
C.~Blake et al., MNRAS \textbf{425}, (2012) 405.

\bibitem{Nadathur2019}
S.~Nadathur, P.~M.~Carter, W.~J.~Percival, H.~A.~Winther and J.~Bautista, Phys. Rev. D \textbf{100}, (2019) 023504.

\bibitem{Chuang2016}
C.-H.~Chuang et al., MNRAS \textbf{461}, (2016) 3781.

\bibitem{Aubert2022}
M.~Aubert et al., MNRAS \textbf{513}, (2022) 186.

\bibitem{Wilson2016}
M.~J.~Wilson, Ph.D. thesis, Edinburgh University, 2017.
    
\bibitem{Zhao2018}
G.-B.~Zhao et al., MNRAS \textbf{482}, (2018) 3497.

\bibitem{Okumura2016}
T.~Okumura et al., PASJ \textbf{68}, (2016) 38.


\end{thebibliography}
\end{document}